\documentclass[final]{statistica}

\usepackage[colorlinks,citecolor=blue,linkcolor=blue]{hyperref}
\usepackage{graphicx}
\usepackage{booktabs} 
\usepackage{amsmath}
\usepackage{multirow}

\title[Analysis of longitudinal data with destructive sampling]{Analysis of longitudinal data with destructive sampling using linear mixed models}
\author{C.A. Avellaneda }%
\email{caavellanedag@unal.edu.co}
\address{Universidad Nacional de Colombia, Bogota, Colombia}
\author{O.O. Melo}%
\email{oomelom@unal.edu.co}
\address{Universidad Nacional de Colombia, Bogota, Colombia}
\author{N.A. Cruz\footnote{Corresponding Author. E-mail: nelson-alirio.cruz@uib.es}}
\email{nelson-alirio.cruz@uib.es}
\address{Universitat de les Illes Balears, Palma de Mallorca, Spain}

\years{202\#}
\volumes{\#}
\anno{LXX\#}
\begin{document}


\section{Introduction}

In many countries, it is common to apply standardized tests that measure secondary education, which must be taken by all students at the end of grade 11 or ISCED level 3 (International Standard Classification of Education \cite{unesco2012international, covacevich2021indicators}) as a prerequisite to obtaining their degree. The particularity of this information is that most students who take this test for the first time do not take it in subsequent years, since the only ones who take it more than once are those who do not pass the eleventh grade. Therefore, we cannot attempt to model the outcomes of each educational institution and each student over several years because it cannot be assumed that the results of a student within the same educational institution are independent between years. The dependency is given because students share an educational institution, teachers, teaching methodologies, and facilities, among other possible factors that can influence learning.

This type of experiment, where the measurement involves the destruction of the unit, that is, a student only presents the test once, is common in other scenarios, such as the quality control of certain devices and the observation of characteristics of fruits, branches or tree leaves that could not be measured at times after their collection or in panel studies that consider different individuals at each time due to difficulties or absence, among others.

In experimental design and linear models, it is of interest to the researcher to detect differences in a variable determined by the study, between groups formed in a particular way, considering one or multiple factors. Furthermore, sometimes the experimental units involved in the study are observed in repeated situations, which constitutes longitudinal data. In general, this means that the usual regression models are not the most appropriate, since their assumptions include independence between measurements, which is violated when observing the same individuals repeatedly.
Various methodologies have been built for modeling longitudinal data that allow modeling the correlation that exists in the same experimental unit, either using linear mixed models or Bayesian mixed models.
In these models, in addition to obtain estimates of the effects of exogenous variables or experimental factors on the response variable, estimates of the behavior of the response variable over time and the correlation between the same measurements of an experimental unit can be constructed \citep{ zou2023bayesian}.

In most models based on longitudinal data, also called panel data, it is assumed that the response variable is measured at each time point for each experimental unit. Suppose there are times when measurements do not exist for some experimental units. In that case, those are modeled as missing data, either by imputing them or using EM algorithms or Bayesian statistics \citep{quintana2016bayesian, ariyo2020bayesian, zou2023bayesian}.
Models have also been built for longitudinal data, where the experimental unit is not measured throughout the entire study but appears in some period, some measurements of it are obtained, and then, it disappears from the study. It is a necessary condition that there are at least two measurements of each experimental unit to be able to estimate their correlation \citep{wan2023multikink, shi2021functional}. For data of this type, \cite{deaton1985panel} proposed a model for a sample of pseudo-panel data, which was constructed with what it called cohorts; These represent groups of individuals with similar characteristics, with the condition that each individual belongs only to a cohort throughout the analysis, and through these groups, the fixed effects corresponding to each of the different cohorts are adjusted.

\citet{gardes2005panel} investigated the biases in the elasticities of income and expenses estimated from methodologies used in different types of samples within which they considered pseudo-panel data, cross-sectional data, and true panels. \citet{verbeek2008pseudo} made an exhaustive review of linear models, their identification, and estimation based on an approach to panel data based on repeated measurements in cross sections, adding lags of the response variable to the explanatory variables as it is usual in time series research methodologies. \citet{sprietsma2012computers} reviewed the influence of access to computing centers and the internet on the effects as pedagogical tools for the acquisition of mathematics skills based on representative samples of school students from the years 1999, 2001, and 2003 in Brazil. The estimation of the model was carried out using the estimator proposed by \citet{deaton1985panel} together with a correction proposed by \citet{verbeek1993minimum} for the case in which there was a small number of measurement times. \citet{tovar2012analysis} considered the case in which the samples selected at different times share some individuals, that is, the different cross sections had people in common, so the fitted model was the one considered by \citet{deaton1985panel} including a temporal autocorrelation in the residuals by having measurements from some individuals, which was done to model the active population in the Basque Country from 1993 to 1999. In the literature, numerous additional cases can be found to those mentioned above where, based on effects associated with fixed values, the behavior of a response variable measured at different times is modeled \citep{urdinola2015long,himaz2016returns,canavire2017non}.

However, none of the models explored in the literature address situations where units observed at one time are considered at later times due to the conditions of the experiment, which is called destructive sampling within the experimental unit. Therefore, in this work a methodology is constructed following the steps described below: i) it is considered that all educational institutions are represented by their students in all measurement periods, ii) each school is defined as an experimental unit, while that the students, from whom the measurements come, are the observation units and are destroyed at the time of measurement and iii) pseudo observational units that are measured throughout the entire study and allow studying the correlation of observational units within an experimental unit is built by grouping students into study levels each year; that is, the students with excellent performance of each year make up a pseudo observational unit, and the same with the students with low or average performance, iv) a mixed effects linear regression model is proposed for the analysis of longitudinal measurements where each unit observation has destructive sampling but comes from experimental units that are maintained from an initial time and throughout the period.
The statistical model constructed is compared with a pseudo-panel linear model that uses averages per school, the usual in these analyses, and a multivariate linear model that assumes independence between all observations adjusted by traditional regression methods. The comparison of the different models used is carried out using the mean square error and confidence intervals, demonstrating the superiority in terms of mean square error and goodness-of-fit of the proposed methodology.

This document is composed as follows: Section 1 describes the statistical methodologies to be used in the development of the proposed model, Section 2 describes the process of making the comparison between methodologies through simulation, and Section 3, an application of these methodologies to a database that contains the scores of the Saber 11 tests carried out by Colombian students from different municipalities during the years 2013-2018, considering that the majority of students who take the exam do not do it again in subsequent years. Finally, the conclusions are presented.

\section{Linear mixed models with destructive sampling}
\subsection{Proposed model}
Linear models are a way of explaining the dispersion of one or more random responses in terms of a series of independent variables, also known as \citet{west2014linear} exogenous. The effects of independent variables can be classified as fixed or random effects. According to \citet{Melo}, when at the end of the experiment conclusions are formulated about a pre-established number of treatments, the model is called fixed effects, so, the statistical inference is made on the average effects of the treatments. If the levels of an attribute come from a random sample from a population of possible selections, the model is called random effects; so, the conclusions are formulated on a greater number of treatments than those used in the experiment, and statistical inference is made about their variances. Models that include fixed and random effects are called mixed effects.
\\
Let $y_{ijk}$ be the response of the $j$-th observational unit carried out at time $k$-th within the $i$-th experimental unit, $i=1, \ldots, n$, $j=1, \ldots,n_i$ and $k=1, \ldots, t$. Therefore, $y_{ijk}$ belongs to repeated measures. In these cases, each experimental unit can have a particular effect on the response variable that is not directly due to the combination of exogenous variables presented. In addition, each observational unit nested within the observational unit can also present effects on the response variable, since the analyses of longitudinal samples usually cover only a sample of units from a larger population and conclusions are desired for the entire set of experimental units. The linear model for this mixed-effects scenario is given by  Equation \eqref{marco:modmix} \citep{bates2000mixed}.
\begin{equation}\label{marco:modmix}
y_{ijk}=\pmb{x}_{ijk}\pmb{\beta}+b_i + \eta_{j(i)} +\epsilon_{ijk}
\end{equation}
where $i=1,\ldots, n$, $j=1, \ldots, n_i$,  $k=1, \ldots, t$,  $\pmb{x}_{ijk}$ is the design matrices of the fixed effects, $\pmb{\beta}$ of dimension $p$ contains the coefficients associated with the fixed effects, while $ \epsilon_{ ijk}$ is the vector of measurement errors within each experimental unit. $b_i$ is the effect of the $i$-th experimental unit on the variable and it is assumed that ${b}_i \overset{\text{iid}}{\sim} {N}(0,\sigma^2_b)$. $\eta_{j(i)}$ is the random effect of the $j$-th observational unit that belongs to the $i$-th experimental unit, assuming $\eta_{j(i)}\overset{\text{iid}}{\sim} N( 0, \sigma^2_{\eta})$, also $\pmb{\epsilon}_{ij}=\{\epsilon_{ij1}, \ldots, \epsilon_{ij t}\} \overset{\text{iid}}{\sim}  {N}_t(\pmb{0}_t,\pmb{\Sigma}_{t \times t})$. In turn, $\epsilon_{ijk}$, $b_i$ and $\eta_{j(i)}$ are assumed independent. The matrix $\pmb{\Sigma}$ associated with the residuals within the measurements of the same unit can consider different situations, that is, it can be associated with autocorrelated processes, for example, those found in the literature related to a series of time, like the \textit{ARIMA} \citep{wei2006time} models, or like those studied in spatial statistics, which are characterized by their variogram function \citep{schabenberger2017statistical}. 
\\
The problem occurs when sampling is destructive with the observational unit, since each experimental unit $j$ is only observed once and never again, so the effect $\eta_{j(i)}$ in Equation \eqref{marco:modmix} is impossible to estimate. However, since within each experimental unit $i$ at time $k$, there are several measurements of experimental units, $b_i$ is not enough to capture the variability between observational units within each experimental unit, and the observational units may not be independent of each other over time, but since they were only measured once, it is not possible to quantify this correlation.  The destructive sampling scheme is shown in Table \ref{simulacion:tabilustracion} to exemplify this idea.
\begin{table}
\centering
\caption{Illustration of the sampling scheme considered in the simulation, each $\ast$ represents the time where the observation is taken from the ``observational units".}\label{simulacion:tabilustracion}
\begin{tabular}{ cc|cccccc } 

\hline\multirow{2}{*}{Experimental unit} & \multirow{2}{*}{Observational unit} & \multicolumn{4}{c}{Time} & \multirow{2}{*}{POU} \\
         & & $T_1$ & $T_2$ &$\hdots$ & $T_t$& \\ \hline
$\text{EU}_1$ & $\text{OU}_1$ & $\ast$ & & & &$\text{PUO}_{1'}$ \\
      $\text{EU}_1$ & $\text{OU}_2$ & $\ast$ & & &&$\text{PUO}_{2'}$ \\
      $\text{EU}_1$ & $\text{OU}_3$ & & $\ast$ & &&$\text{PUO}_{1'}$ \\
      $\text{EU}_1$ & $\text{OU}_4$ & & $\ast$ & &&$\text{PUO}_{1'}$ \\
$\vdots$ & $\vdots$ & & & $\ddots$ &&$\vdots$ \\
$\text{EU}_1$ & $\text{OU}_{n_1-1}$ & & & &$\ast$&$\text{PUO}_{2'}$ \\
         $\text{EU}_1$ & $\text{OU}_{n_1}$ & & & & $\ast$ &$\text{PUO}_{2'}$\\ \hline
$\text{EU}_2$ & $\text{OU}_1$ & $\ast$ & & &&$\text{PUO}_{1}$ \\
$\text{EU}_2$ & $\text{OU}_2$ & $\ast$ & & &&$\text{PUO}_{2}$\\
$\text{EU}_2$ & $\text{OU}_3$ & & $\ast$ & & &$\text{PUO}_{1}$\\
$\text{EU}_2$ & $\text{OU}_4$ & & $\ast$ & & &$\text{PUO}_{2}$\\
$\vdots$ & $\vdots$ & & & $\ddots$ &&$\vdots$ \\
$\text{EU}_2$ & $\text{OU}_{n_2-1}$ & & & & $\ast$&$\text{PUO}_{1}$ \\
$\text{EU}_2$ & $\text{OU}_{n_2}$ & & & & $\ast$&$\text{PUO}_{2}$ \\
		\hline

\end{tabular}
\end{table}
Therefore, in this work, a model is proposed that forms groupings of observation units with a similar response to another observational unit at another instant of time within the same experimental unit. This will guarantee that each experimental unit has measurements of these observation units at all times. Therefore, the model given in Equation \eqref{marco:modmix} will be:
\begin{equation}\label{marco:modmix1}
y_{irk}=\pmb{x}_{irk}\pmb{\beta}+b_i + \eta_{r(i)} +\epsilon_{irk}, \enspace
\end{equation}
with $i=1,\ldots, n$, $r=1, \ldots, G_i$, $k=1, \ldots, t$, where the only change with respect to Equation \eqref{marco:modmix} is that each $j$ is only observed in an $ik$-th combination. The value of $r$ contains observational units $j$ from the first period $(k=1)$, second period $(k=2) $, and so on over all values of $k$ for each value of $i$. The total of groups proposed within each experimental unit is $G_i$. An example of these pseudo-observational units is shown in the $POU$ column of Table \ref{simulacion:tabilustracion}, where two of these are constructed, $1'$ and $2'$ by ordering the response variable in ascending order each time to capture the correlation.
In the context of standardized tests in schools, those students with good performance are equated at each of the moments within each experimental unit. Likewise, the same is done with those with poor performance. In each of the classrooms, two subgroups can be formed: those with good and poor performance, which could be monitored over time. The above process can be carried out even when the students are different at different measurement moments.
In Table \ref{simulacion:tabilustracion}, observational units that make up the group with good performance at time 1 are different, due to destructive sampling, from the individuals with the same performance in the same experimental unit but at different times. However, it is assumed that there is dependence because they are immersed in the same experimental unit. This allows the construction of longitudinal observation for the study.  Finally, it is assumed that the random effects are independent of each other and that each of them follows a normal distribution, as follows:
\begin{equation}\nonumber
 b_{i} \overset{\text{iid}}{\sim} N \left( 0,\sigma _{b}^2 \right), \enspace \eta_{r(i)}\overset{\text{iid}}{\sim} N \left( 0,\sigma _{\eta}^2 \right) \enspace \text{and} \enspace \pmb{\epsilon}_{ir}=(\epsilon_{ir1}, \ldots,\epsilon_{irt})\overset{\text{iid}}{\sim} N \left( 0,\pmb{\Sigma} _{t\times t} \right).
\end{equation}

\subsection{Estimation}
For the estimation, it is noted that the model given in Equation \eqref{marco:modmix1} can be written as:
\begin{equation}\label{marco:modmix2}
\pmb{y}_i=\pmb{X}_i\pmb{\beta}+\pmb{Z}_{i}{\pmb{v}_{i}}+\tilde{\pmb{\epsilon}}_i
\end{equation}
where $\pmb{y}_i$ contains all the observations of the experimental unit $i$, $\pmb{X}_i$ is the design matrix of the fixed effects associated with the parameter vector $\pmb{\beta}$, ${\pmb{v}_{i}}=(b_i, \eta_{1(i)}, \ldots, \eta_{G_i(i)})^t$ and the matrix $\pmb{Z}_{i}$ is the design matrix of them, and $\tilde{\pmb{\epsilon}}_i$ is the random error of the experimental unit $i$.  From the assumptions made on the model given in \eqref{marco:modmix}, it is obtained that:
\begin{align} 
\pmb{y}_i|{\pmb{v}_{i}} &\sim \pmb{N}(\pmb{X}_i \pmb{\beta}+\pmb{Z}_i{\pmb{v}_{i}},\sigma^2 \pmb{I}), \enspace {\pmb{v}_{i}} \sim \pmb{N}(\pmb{0},\pmb{\Psi}). \nonumber 
\end{align}
The likelihood function for this model adapting the methodology of \citet[p. 64]{bates2000mixed} is as follows: \begin{equation}\label{marco:memlik} 
\nonumber L(\pmb{\beta},\pmb{\theta},\sigma^2) = \frac{\text{exp} \left \{\sum _{i=1}^n \|\pmb{\tilde{y}}_i-\pmb{\tilde{X}}_i\pmb{\beta} -\pmb{\tilde{Z}}_i\pmb{\hat{v}}_i\|/2\sigma^2 \right\}}{\left(2 \pi \sigma^2 \right)^{nt/2}} \prod_{i=1}^n \frac{|\pmb{\Delta}|}{\sqrt{| \pmb{\tilde{Z}}_i^t \pmb{\tilde{Z}}_i |}} 
\end{equation}
where 
\begin{align}
\nonumber
    \pmb{\tilde{y}}_i&= \begin{bmatrix} \pmb{y}_i \\
\pmb{0}_{q\times q}
\end{bmatrix}, \enspace 
\pmb{\tilde{X}}_i= \begin{bmatrix} \pmb{X}_i \\
\pmb{0}_{q\times q}\end{bmatrix},
\enspace \sigma^2 \pmb{\Psi}^{-1}=\pmb{\Delta}^t\pmb{\Delta}, \enspace
\text{, and} \enspace 
\pmb{\tilde{Z}}_i= \begin{bmatrix} \pmb{Z}_i \\ \pmb{\Delta} 
\end{bmatrix}.    \\
     \pmb{\hat{v}}_i&= \left( \pmb{\tilde{Z}}_i^t \pmb{\tilde{Z}}_i \right)^{-1} \pmb{\tilde{Z}}_i^t \left( \pmb{\tilde{y}}_i-\pmb{\tilde{X}}_i\pmb{\beta} \right). \label{marco:b}
\end{align}
The restricted maximum likelihood method (REML) is the alternative estimation method by which unbiased estimators for the variance components will be obtained and which will be carried out. The process of deducing the function to be optimized for this methodology is adapted from \citet[p. 75]{bates2000mixed} and as in the case of maximum likelihood, a function to be optimized is obtained by iterative methods for a group of parameters.
For the methodology described in this section, the R software \citep{R} has two main libraries \textit{``lme4"} and \textit{``nlme"} (see \citet{lme4} and \citet{nlme}, respectively) that allow carrying out the adjustments, estimation and evaluation of the described models. \citet[p. 90]{finch2016multilevel} explains the use of the main functions of these two packages, which are $lme(\cdot)$ for \textit{``nlme"} and $lmer(\cdot)$ for \textit{ ``lme4"}, that is,  in addition to the different arguments and functions that complement their use, a description of the expressions required for the formulas used to include fixed and random factors through pre-established databases. The evaluation of the assumptions required by the model in the software is described in \citet[p. 174]{bates2000mixed}.
The R code for this estimation is presented in the supplementary file \ref{sf1}.

\subsection{Comparison}

Comparing the estimators with the grouping of observational units obtained with the model proposed in \eqref{marco:modmix1} with the following: i) obtained by the \cite{deaton1985panel} model, which averages the observational units for each experimental unit within each period and takes into account a random effect of each experimental unit, and it is defined by:
\begin{equation}\label{marco:modmix10}
\bar{y}_{i\cdot k}=\bar{\pmb{x}}_{i\cdot k}\pmb{\beta}+b_i^{(i)} +\bar{\epsilon}^{(i)}_{i\cdot k}, \enspace
\end{equation}
ii) an exclusive model of fixed effects defined by:
\begin{equation}\label{marco:modmix11}
y_{ijk}=\pmb{x}_{ij k}\pmb{\beta}+\epsilon^{(ii)}_{ijk}, \enspace
\end{equation}
and iii) a model with random effect of each experimental unit and the fixed effects similar to i) defined by:
\begin{equation}\label{marco:modmix12}
y_{ijk}={\pmb{x}}_{ij k}\pmb{\beta}+b_i^{(ii)} +\epsilon^{(iii)}_{ijk}, \enspace
\end{equation}
then, the following result is obtained:
\begin{lem}\label{lem1}
Assuming that the model proposed in Equation \eqref{marco:modmix} is true and that each observational unit $k$ is measured destructively, and $\pmb{X}$ is conformed only with dichotomous variables, let 
$$MSE^{(0)} = \frac{1}{\sum_i n_i}\sum_{ijk}\left(y_{ijk}-\hat{y}^{(0)}_{ijk}\right)^2$$
for the model proposed in Equation \eqref{marco:modmix1}, and analogously for the other models defined in equations \eqref{marco:modmix10}, \eqref{marco:modmix11} and \eqref{marco:modmix12}, then it holds that:
        $$MSE^{(0)} \leq MSE^{(iii)} \leq MSE^{(i)}\leq MSE^{(ii)}$$
\end{lem}
\begin{proof}
See \ref{apenA}
\end{proof}
Lemma \ref{lem1} guarantees that the proposed model is better than the usual models in terms of MSE. However, in the following simulation study, the test statistics for hypothesis contrast will be compared to demonstrate the robustness of the proposal. 

\section{Simulation}
To carry out the comparison between the proposal presented in the previous section and different types of usual regression models for the destructive sampling scheme of observational units, a simulation exercise will be carried out. In this study, the average square of the errors of the estimates of each model will be used, that is, its mean square error ($MSE$), as a way to make the comparison between the different methodologies described.
Table \ref{simulacion:tabilustracion} shows the sampling scheme with two experimental units and from each of them two observational units measured at each of the $t$ times, that is, the symbol $\ast$ corresponds to the moment where the measurement is made on the units. There, it can be seen that the observational units measured at one time are not measured later.

Without loss of generality, for simplicity in this exercise, a single treatment with two groups will be taken into account, whose effects on the simulated response will be denoted as $A_m$ with $m=1,2$. On the other hand, the random effects considered are $b_{i}$, each one is the effect of the $i$-th experimental unit on the treatment $m$, $\eta_{j(i)}$ is the effect of the $j$-th observational unit within the $i$-th experimental unit of the $m$-th treatment, and finally, $\epsilon_{ijkm}$ is a random error associated with the measurements within the ``observational units".

In this way, the statistical model adapted from Equation \eqref{marco:modmix1} for the simulation is
\begin{equation}\label{simulacion:eqmodsimula}
     y_{ijklm} =\mu + A_{m}+b_{i}+\eta_{j(i)}+T_k+AT_{mk}+\epsilon_{ijklm}
     \end{equation}
with $i=1,\ldots, n$, $j=1, \ldots, J$, $ k=1,\ldots,t$, $l=1, \ldots, L$, and $n$ is the number of experimental units in each treatment, $m=1,2$, $n_{im}$ is the number of originally simulated observational units within each experimental unit, $L$ is the number of replicates of each observational unit within each time, $t$ is the number of times considered and $\mu$ is a global average. On the other hand, $A_m$ and $T_k$ are the effects of the $m$-th treatment and the $k$-th time, respectively, and $AT_{mk}$ is the effect of their interaction. Finally, $y_{ijklm}$ is the response variable measured in the $j$-th observational unit of the $i$-th experimental unit at the $k$-th time and corresponding to the $m$-th treatment. This type of notation can be consulted in detail in \citet{Melo}.
Due to estimability conditions, the following restrictions are assumed in this model:
\begin{equation}\label{estimabilidad}
  \sum _{m=1}^2 A_m= \sum _{k=1}^t T_k=\sum _{m=1}^2 AT_{mk} = \sum _{k=1}^t AT_{mk}= 0
\end{equation}
\noindent Additionaly, $b_{i} \sim N(0,\sigma_b^2)$, $\eta_{j(i)}\sim N(0,\sigma_{\eta}^2)$ and
$\pmb{\epsilon}_{ijm} =(\epsilon_{ij1m}, \epsilon_{ij2m}, \ldots, \epsilon_{ijtm}) \overset{iid}{ \sim} N(0,\pmb{\Sigma})$ which are assumed to be independent of each other $\pmb{\Sigma}=\{\sigma_{ij}\}_{t\times t}=\rho^{|i-j|}$ is the notation of a stochastic autoregressive process of order one.

The first part of the exercise focuses on simulating $b_{i}$, $\eta_{j(i)}$, and $\epsilon_{ijklm}$ for obtaining the response variable $y_{ijklm}$ using  Equation \eqref{simulacion:eqmodsimula}. The second part corresponds to simulating the loss of information, in such a way that a structure similar to that described by Table \ref{simulacion:tabilustracion} is reached. Once there is a single measurement of each of the observational units, the main problem is that an individual follow-up over time could not be done, so a random source of error would be unknown, that is, it would not be possible to obtain an estimate of the effects $\eta_{j(i)}$ and one of its possible consequences would be on the test statistics, affecting the values $p$ and standard errors corresponding to the different coefficients.

\begin{figure}[h!]
    \includegraphics[width=12cm]{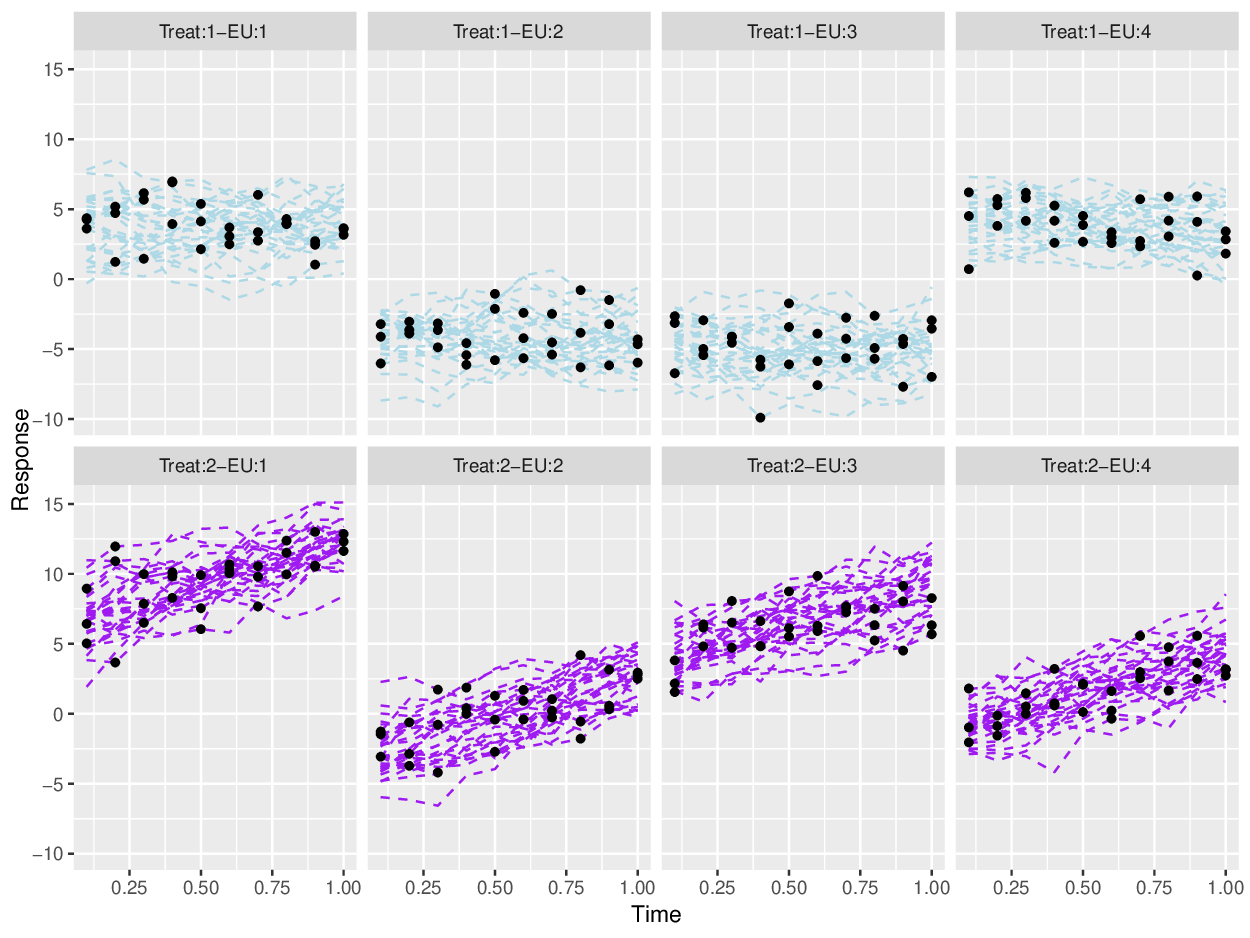}
    \caption{Scatter diagram that illustrates the sampling scheme used, whose dotted lines represent the complete profiles for each observational unit, and the points are the observations selected at each time.}\label{simulacion:fig1}
\end{figure}

As an illustration, a data table associated with the model given in \eqref{simulacion:eqmodsimula} was initially simulated. Figure \ref{simulacion:fig1} shows this procedure, where the dotted lines refer to the time profiles of each observational unit. The colors of the lines identify the assigned treatment and the titles in each box correspond to the numbering of the experimental units and their assigned treatments, while the points are the observations. This means that the simulated values of $y_{ijkm}$ are illustrated on the ordinate axis, while the measurement times are represented on the abscissa axis. In each experimental unit and for each time, four observational units are selected, which are not considered again at later times.

To adjust a statistical model under the considered scheme and simulation, the following models are proposed:
\subsection{Exclusive fixed effects model}\label{simulacion:sec_fijos}
Considering that the observational units are only measured on one occasion, the first approach is the linear regression model that assumes independence between all its measurements in the response variable, only considering a random error in addition to the fixed effects. This model described in equation \eqref{marco:modmix11} is shown in the following Equation:
\begin{gather}\label{simulacion:eqmodsimula_2}
     y_{ijklm} =\mu + A_{m}+T_k+AT_{mk}+\epsilon^{(ii)}_{ijklm}, 
     \end{gather}

\noindent with $i=1,\ldots, n$, $j=1, \ldots, J$, $k=1,\ldots,t$, $l=1,\ldots,L$, $m= 1,\ldots, M$, and also, under the estimability conditions given in  Equation \eqref{estimabilidad}.
The hypotheses that can be evaluated from the variance analysis table of the model given in  Equation \eqref{simulacion:eqmodsimula} and that will be considered in the same way in the following cases are those that are related to the $AT_{mk}$, $A_m$, $T_k$, meaning that it is of interest to evaluate whether there are significant differences between their respective levels. In this way, the null hypotheses for each case:
  \begin{center}
   $H_0^{(1)}$: $AT_{mk}=AT_{m'k'}$ with $m \neq m'$ and $k \neq k'$.\\
   $H_0^{(2)}$: $A_m=A_{m'}$ with $m \neq m'$.\\
   $H_0^{(3)}$: $T_k=T_k'$ with $k \neq k'$.
  \end{center} 
Table \ref{table:fixed_mod} corresponds to the variance analysis of the model of Equation \eqref{simulacion:eqmodsimula_2} with $$SSAT=\displaystyle \sum _{mk} \frac{y_{\cdot \cdot k \cdot m}^2}{nJL}-\sum _{k} \frac{y_{\cdot  \cdot k \cdot \cdot}^2}{MnJL} - \sum _{m} \frac{y_{\cdot \cdot \cdot \cdot m}^2}{tnJL} + \frac{y_{\ldots \cdot \cdot}^2}{MntJL}$$
where only one source of random variation is included, denoted by $\epsilon_{ijklm}^{(ii)}$. The response variable is denoted by $y_{ijklm}$ which corresponds to the measurement of the $l$-th replication of the $j$-th observational units after the selection of destructive sampling, in the $i$-th experimental unit, $k$-th time and  $m$-th treatment. The interpretation of the coefficients $A_m$, $T_k$ and $AT_{mk}$ is exactly the same as that of the model described in Equation \eqref{simulacion:eqmodsimula}.
  
For these purposes, the mean squares of the factors are calculated by dividing the respective sums of squares by their degrees of freedom. The test statistics for the effects in  Equation \eqref{simulacion:eqmodsimula_2} are calculated by dividing the respective mean squares associated with the factors by the mean square of the error, that is, the test statistics related to $H_0^{(1 )}$, $H_0^{(2)}$ and $H_0^{(3)}$ are calculated as follows (see Table \ref{simulacion:eqmodsimula_2}):
 
  \begin{gather}\nonumber
     F_{AT}=\frac{MS(AT)}{MSE}, \enspace F_{A}=\frac{MS(A)}{MSE} \text{ and} \enspace F_{T}=\frac{MS(T) }{MSE}, \text{respectively}.
\end{gather}
 \begin{table}
  \footnotesize
  \centering
\caption{Variance analysis table for the model given in Equation \eqref{simulacion:eqmodsimula_2} }
  \begin{tabular}{cllll} \hline
  Source &
  df &
  Sum of squares &
  E(MSE) & F \\
  \hline
  $A_m$ & $M-1$ & $\displaystyle \sum _{m} \frac{y_{\cdots \cdot m}^2}{ntJL}- \frac{y_{\cdots \cdot \cdot }^2 }{MntJL}$ & $\displaystyle \sigma _{\epsilon^{(ii)}}^2+\frac{ntJL}{(M-1)}\sum _{m}A_{m}^2$ &$F_A$ \\
  $T_k$ & $t-1$ & $\displaystyle \sum _{k} \frac{y_{j \cdot \cdot \cdot \cdot}^2}{MnJL}- \frac{y_{\cdots \cdot \cdot}^2}{MntJL}$ & $ \displaystyle \sigma _{\epsilon^{(ii)}}^2+\frac{MnJL}{(t-1)}\sum _{k}T_{k}^2$ &$ F_T$\\
  $AT_{mk}$ & $(M-1)(t-1)$ & $SSAT$ & $\displaystyle \sigma_{\epsilon^{(ii)}}^2+\frac{nJL}{(M-1)( t-1)}\sum_{mk}AT_{mk}^2$ &$ F_{AT}$\\
  $\epsilon_{ijklm}^{(ii)}$ & $Mt(nJL-1)$ & $ \displaystyle \sum _{ijklm} y_{ijklm}^2 - \sum_{mk}\frac{y_{j \cdot  \cdot \cdot m}^2}{nJL}$ & $\displaystyle \sigma _{\epsilon^{(ii)}}^2$ &\\ \hline
  Total & $MntJL-1$ & $ \displaystyle \sum _{ijklm} y_{ijklm}^2 - \frac{y_{\cdots  \cdot \cdot }^2}{MntJL}$ && \\
  \hline
\end{tabular}
\label{table:fixed_mod}
\end{table}

\subsection{Application of the model for pseudo-panel}\label{simulation:sec_deaton}

In \citet{deaton1985panel} a regression model is proposed for the case in which measurements are made periodically to groups of individuals, among which there is a rotation, that is, the people who are observed at a particular time can or not present for subsequent measurements for reasons external to the research.
In the context of simulation, considering that the observational units are measured at a single time due to experimental conditions, a model can be applied for pseudo-panel data, such as the one described in  Equation \eqref{simulacion:eqmodsimula_3}, where the groupings will be made up of the experimental units since they are considered at all times in the study. For this work, the factor associated with the experimental units, which is included in the regression model, is random, since we are only working with a random sample of these units, and the conclusions obtained are intended to be generalized to the entire population.
For this model, adapting Equation \eqref{marco:modmix10} the averages are calculated per experimental unit in each time from Equation \eqref{simulacion:eqmodsimula} obtaining:
\begin{align}\label{simulacion:eqmodsimula_3}
\bar{y}_{i\cdot k \cdot m} &= \sum _{jl} \left \{\mu + A_{m}+b_{i}+\eta _{r(i)}+T_k+AT_{mk}+ \epsilon _{ijklm} \right \}/JL \\ \nonumber
&= \mu + A_{m}+b_{i}^{(i)}+T_k+AT_{mk}+\bar{\epsilon}_{i\cdot k \cdot m} \nonumber    
\end{align}

\noindent with $i=1,..., n$, $j=1,\ldots,J,$, $k=1,2,\ldots,t$ and $l=1,2,\ldots, L$, where $b_{i}^{(i)}= \sum_{jl}\frac{\left(b_{i}+\eta_{r(i)} \right)}{JL}$, $\bar{y}_{ i\cdot k \cdot m}=\sum _{jl}\frac{y_{ijklm}}{JL}$ and $\bar{\epsilon}_{i\cdot k \cdot m}=\sum _{jl}\frac {\epsilon _{ijklm}}{JL}$ and with the estimability conditions defined in \eqref{estimabilidad}. On the random factors, it is assumed that $b_{i}^{(i)}\sim N \left( 0,\sigma _{b^{(i)}}^2 \right) \enspace \text{and} \enspace \bar{ \epsilon}_{ij.m}\overset{\text{iid}}{\sim} N \left( 0,\sigma _{\bar{\epsilon}}^2 \right).$ For the model given in Equation \eqref{simulacion:eqmodsimula_3}, $\bar{y}_{i\cdot k \cdot m}$ and $\bar{\epsilon}_{i\cdot k \cdot m}$ are the response variable and the average random error for the $i$-th experimental units in the $j$-th time allocated in the $m$-th treatment, respectively, while $b_{i}^{(i)}$ is the random effect associated with the $i$-th experimental units in the $m$-th treatment. The interpretation of the coefficients $A_i$, $T_k$, and $AT_{mk}$ is the same as that of the model described in Equation \eqref{simulacion:eqmodsimula}.
\begin{table}
  \footnotesize
  \centering
  \caption{Variance analysis table for the model given in Equation \eqref{simulacion:eqmodsimula_3}.}
   \begin{tabular}{cllll} \hline
  Source &
  df &
  Sum of squares &
  E(MSE) &F \\ \hline
  $A_m$ & $M-1$ & $\displaystyle \sum _{m} \frac{y_{\cdot \cdot \cdot \cdot m}^2}{tn}- \frac{y_{\cdots \cdot } ^2}{Mnt}$ & $\displaystyle \sigma _{\epsilon}^2+t\sigma _{b^{(i)}}^2+\frac{tn}{(M-1)}\sum _{m}A_{ m}^2$ &$F_A$ \\
 
  $b_{i}^{(i)}$ & $M(n-1)$ & $\displaystyle \sum _{ijm} \frac{y_{\cdot k m}^2}{t}- \frac{y_{ \cdots m }^2}{tn}$ & $\displaystyle \sigma _{\epsilon}^2+t\sigma _{b^{(i)}}^2$ &$F_b$\\
 
  $T_k$ & $t-1$ & $\displaystyle \sum _{k} \frac{y_{\cdot j\cdot \cdot \cdot}^2}{Mn}- \frac{y_{\cdots \cdot \cdot}^2}{Mnt}$ & $ \displaystyle \sigma _{\epsilon}^2+\frac{Mn}{(t-1)}\sum _{k}T_{k}^2$ &$F_T$\\
 
  $AT_{mk}$ & $(M-1)(t-1)$ & $SSAT$ & $\displaystyle \sigma_{\epsilon}^2+\frac{n}{(M-1)(t- 1)}\sum_{mk}AT_{mk}^2$&$F_{AT}$ \\
 
  $\epsilon_{ij\cdot m}$ & $M(t-1)(n-1)$ & $ SSE$ & $\displaystyle \sigma _{\epsilon}^2$ &\\ \hline
  Total & $Mnt-1$ & $ \displaystyle \sum _{ijm} y_{ij\cdot \cdot m}^2 - \frac{y_{\cdots  \cdot \cdot }^2}{Mnt}$  && \\
  \hline
\end{tabular}
\label{table:mod_deaton}
\end{table}
From the model given in Equation \eqref{simulacion:eqmodsimula_3}, its variance analysis table given in Table \ref{table:mod_deaton} is deduced, where a random effect associated with the experimental units measured in the time, for Table \ref{table:fixed_mod} of the model in Equation \eqref{simulacion:eqmodsimula_2} and the values of $SS(AT)$ and $SSE$ are given by the following expressions:
\begin{align*}
     SS(AT) &= \sum _{mk} \frac{y_{ \cdot \cdot k\cdot m}^2}{n}-\sum _{k} \frac{y_{ \cdot \cdot k \cdot \cdot}^ 2}{Mn} - \sum _{m} \frac{y_{\cdot \cdot \cdot \cdot m}^2}{tn} + \frac{y_{\cdots\cdot \cdot}^2}{Mnt }\\
     SSE &= \sum _{ikm} y_{i\cdot k\cdot m}^2 - \sum _{mk}\frac{y_{ \cdot \cdot k\cdot m}^2}{n}-
  \sum_{im}\frac{y_{i\cdot \cdot \cdot m}^2}{t}+
   \sum _{m}\frac{y_{\cdot \cdot\cdot \cdot m}^2}{tn}
\end{align*}
The inclusion of this random effect causes the $MSE$ of Tables \ref{table:fixed_mod} and \ref{table:mod_deaton} to be different; therefore, the test statistics and conclusions may be different. To evaluate $H_0^{(1)}$, $H_0^{(2)}$ and $H_0^{(3)}$, test statistics are calculated, respectively:

\begin{gather}\nonumber
     F_{AT}=\frac{MS(AT)}{MSE}, \enspace F_{T}=\frac{MS(T)}{MSE} \text{and} \enspace F_{A}=\frac{ MS(A)}{MS(b^*)}.
\end{gather}

The model given in Equation \eqref{simulacion:eqmodsimula_3} is known in the experimental design literature as a specific case of divided plots. All the details and a great variety of this type of design can be consulted in \citet{federer2007variations}. Comparing this model with the original regression described in Equation \eqref{simulacion:eqmodsimula}, the absence of the random effect associated with the observational units measured over time denoted as $\eta_{j(i)}$ is observed, which done because the impact of the absence of this factor must evaluate.

\subsection{Proposed mixed effects model}\label{simulation:sec_proposed}

With the model given in Equation \eqref{simulacion:eqmodsimula_3}, an average effect of the experimental units can be adjusted but in comparison with the original simulated model of Equation \eqref{simulacion:eqmodsimula}, a random effect may be omitted associated with the observational units, denoted by $\eta_{r(i)}$, which can be modeled by Equation \eqref{marco:modmix1}, that is, the one proposed in this paper. It is clear that each unit is measured only once and with a single measurement, it is not feasible to estimate a parameter. In this way, the proposed is to form subgroups of observational units within each experimental unit, so that the effect of each of these subgroups can be estimated for the response variable, which is done not to omit sources of variation. It can be thought that within each classroom different groups will be formed according to their gender, date of birth, or some characteristic of the students. Applying this idea to the results illustrated in Figure \ref{simulacion:fig1}, within each experimental unit, two groups of observations are obtained ($G=2$). It should be noted that an initial approach is with two groups within each experimental unit, but the idea can be worked with more groups, in which case they could be formed by divisions of the smaller values such as quartiles, certain percentiles, or from auxiliary information of the observational units that have not been included within the statistical model, such as the gender of each student for the aforementioned example.
In this way, the model considered is given in Equation \eqref{simulacion:eqmodsimula_4}.
\begin{gather} \label{simulacion:eqmodsimula_4}
y_{irklm} = \mu + A_{m}+b_{i}+\eta_{r(i)}+T_k+AT_{mk}+\epsilon_{ijklm}
\end{gather}

\noindent with $i=1,\ldots, n$,  $r=1,2$, $k=1,2,\ldots,t$, $l=1,2$,
$m=1,\ldots,M$. The estimability conditions are those given in \eqref{estimabilidad}, $b_{i}$ is the random effect associated with the experimental units, while $\eta_{r(i)}$ is the effect associated with the groupings formed within each experimental unit. In this model, $y_{ijklm}$ is the response associated with the $l$-th replication in the $k$-th group of observational units of the $i$-th experimental units for the $j$-th time and in the $m$-th treatment and $\epsilon_{ijklm}$ is its respective residual. Table \ref{table:proposed_mod} shows the analysis of variance corresponding to the model in question, where:
   \begin{align*}
       SS(AT) &=\sum _{ik} \frac{y_{i \cdot k \cdot \cdot}^2}{JL}-\sum _{k} \frac{y_{ \cdot \cdot k \cdot\cdot}^2 }{MnJL} - \sum _{m} \frac{y_{\cdot \cdot\cdot \cdot m}^2}{tnJL} + \frac{y_{\cdots \cdot \cdot}^2}{MntJL} \\
       SSE & = \sum _{ijklm} y_{ijklm}^2 - \sum _{mk}\frac{y_{\cdot \cdot k \cdot m}^2}{nJL}-
  \sum_{iklm}\frac{y_{i\cdot j \cdot}^2}{t}+
   \sum _{m}\frac{y_{\cdot \cdot \cdot \cdot m}^2}{tnJL}
   \end{align*}
Finally, it is assumed that the random effects are independent of each other, and  each of them follows a normal distribution, as follows:
 \begin{center}
 $b_{i}\sim N \left( 0,\sigma_{b}^2 \right), \enspace \eta_{r(i)}\overset{\text{iid}}{\sim}N \left( 0,\sigma_{\eta'}^2 \right) \enspace  \text{and} \enspace \epsilon_{ijklm}\overset{\text{iid}}{\sim} N \left( 0,\sigma_{\epsilon}^2 \right).$ \end{center}

\begin{table}
  \footnotesize
  \centering
  \caption{Variance analysis table for the model given in Equation \eqref{simulacion:eqmodsimula_4}.}
  \begin{tabular}{clll} \hline
  Source &
  df &
  Sum of squares &
  E(MSE)\\ \hline
  $A_i$ & $M-1$ & $\displaystyle \sum _{m} \frac{y_{\cdot\cdot \cdot \cdot m}^2}{ntJL}- \frac{y_{\cdots \cdot\ cdot}^2}{MntJL}$ & $\displaystyle \sigma _{\epsilon}^2+t\sigma_{\eta'}^2+tJL\sigma _{b}^2+\frac{ntJL}{( M-1)}\sum_{m}A_{m}^2$ \\
 
  $b_{i}$ & $M(n-1)$ & $\displaystyle \sum _{ijm} \frac{y_{ij\cdot \cdot m}^2}{t}- \frac{y_{m \cdots}^2}{ntJL}$ & $\displaystyle \sigma _{\epsilon}^2+t\sigma_{\eta'}^2+tJL\sigma _{b}^2$ \\
 
  $\eta_{r(i)}$ & $nM(JL-1)$ & $\displaystyle \sum _{ijkm} \frac{y_{i\cdot \cdot\cdot m}^2}{t}- \frac {y_{i\cdot \cdot \cdot m }^2}{tJL}$ & $\displaystyle \sigma_{\epsilon}^2+t\sigma_{\eta'}^2$ \\
 
  $T_k$ & $t-1$ & $\displaystyle \sum _{k} \frac{y_{ \cdot \cdot j \cdot \cdot}^2}{MnJL}- \frac{y_{\cdots \cdot \cdot}^2}{MntJL}$ & $ \displaystyle \sigma _{\epsilon}^2+\frac{MnJL}{(t-1)}\sum _{k}T_{k}^2$ \\
 
  $AT_{mk}$ & $(M-1)(t-1)$ & $SSAT $ & $\displaystyle \sigma_{\epsilon}^2+\frac{nJL}{(M-1)(t- 1)}\sum_{mk}AT_{mk}^2$ \\
 
  $\epsilon _{ijklm}$ & $M(t-1)(nJL-1)$ & $ SSE$ & $\displaystyle \sigma _{\epsilon}^2$ \\ \hline
  Total & $MntJL-1$ &  $ \displaystyle \sum _{ijklm} y_{ijkl m}^2 - \frac{y_{\cdots  \cdot \cdot }^2}{MntJL}$& \\
  \hline
\end{tabular}
\label{table:proposed_mod}
\end{table}

\normalsize

\subsection{Multivariate analysis of variance model}\label{simulation:sec_manova}

A final model considered is the multivariate analysis of variance model. Based on the sampling scheme, it is not possible to individually monitor the observational units to form the response variable vectors; it is made of the groupings made in the Section \ref{simulation:sec_proposed}, where two groups were formed within each experimental unit. With these, it is to calculate the average in each of them, that is,
 
  \begin{gather}
   \bar{y}_{ijk \cdot m}=\sum_l \frac{y_{ijklm}}{L}, \nonumber
  \end{gather}
 
   \noindent which is the average of the $k$-th cluster of the $i$-th experimental unit at the $j$-th time. In this way, the response vectors over time are given in the form
 
\begin{center}
     $\pmb{y}_{ikm}=\left[\bar{y}_{i1k\cdot m} \enspace \bar{y}_{i2k\cdot m} \enspace \cdots \enspace \bar{y }_{itk\cdot m} \right]$.
\end{center}

\subsection{Simulation results}

The built code considered a destructive sampling scheme, that is, once the model given in Equation \eqref{simulacion:eqmodsimula} was simulated, a random sample of observational units was taken, as previously described. To obtain more accurate conclusions each time the data was simulated one hundred times, obtaining a final sample as illustrated by the points of Figure\ref{simulacion:fig1} and the adjustment of the four models considered. The R code for this simulation is presented in supplementary file \ref{sf1}.

Based on the four models described, the respective evaluation was carried out to determine which was the best. Since the first three approaches in sections \ref{simulacion:sec_fijos}, \ref{simulation:sec_deaton} and \ref{simulation:sec_proposed} consider the same fixed effects, the estimates of the coefficients $A_i$, $T_k$, $AT_{ij}$ are the same, but what changes are the estimates of their standard errors, and in general, the mean square error with which the test statistics are calculated. It is because the comparison was made based on the $MSE$ and not with respect to the estimates of these parameters.

To simulate the model given in Equation \eqref{simulacion:eqmodsimula}, $t=10$, $K=4$, $L=10$ was considered, and in addition, the following values were taken for the variances:
\begin{center}
$\sigma_b^2=5$, $\sigma^2_\eta=4$ and $\sigma^2_\epsilon=2$.
\end{center}
In the first instance, the values of $\epsilon_{ijJL}$ were simulated from an autoregressive process of order one ($AR(1)$) with a parameter $\rho=0.8$. A difference between treatments $|A_2-A_1|=A \in \{0, 0.1, 0.4, 0.6, 0.8, 1\}$ was taken, an increase in time of approximately five units in the response variable for each unit of increase in time. Finally, an interaction between the time and treatment factors was included, in which several cases were considered, that is, different scenarios were simulated depending on a maximum difference between the levels of said interaction denoted as $\Delta AT_{max} $, which takes values from zero to one.

\begin{figure}
    \includegraphics[width=12cm]{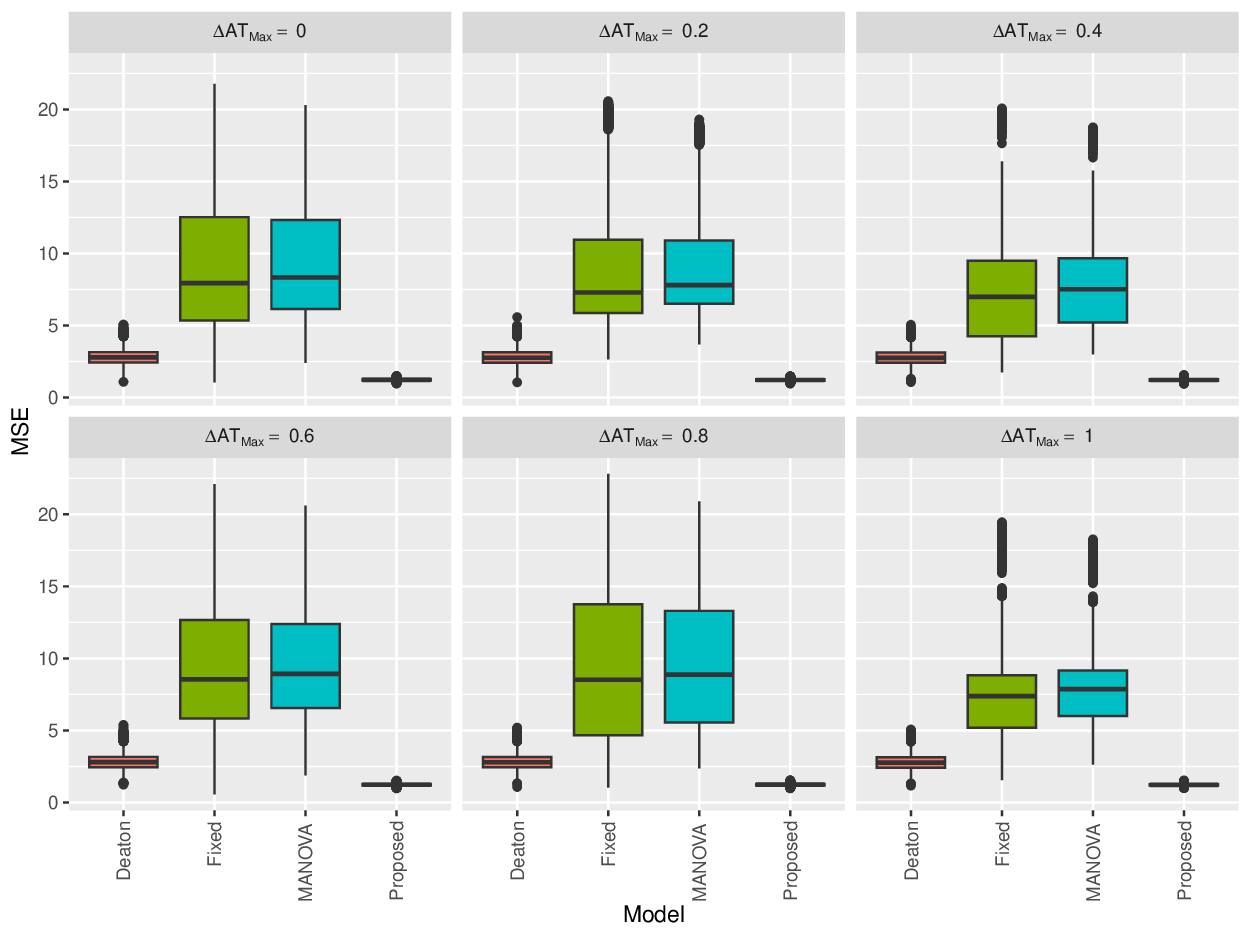}
    \caption{$MSE$ for 100 simulations of the destructive sampling scheme for the four models.}\label{simulation:fig4}
\end{figure}

In the first part of the comparison, a general model was simulated (without loss of information) based on the given coefficients, and for this, the destructive sampling of observational units was generated. In such way, there is only one measurement of each one at a specific time (incomplete data). From these, the four models described were adjusted and the $MSE$ was calculated for each of them. The previous procedure was repeated 100 times, so that for each case four $MSE$ were collected, one for each model. Figure \ref{simulation:fig4} shows the distributions of the 100 values per group, using boxplots. The headings of each of the boxes correspond to the values taken by $\Delta AT_{max}$. There, it is observed that the model with the highest values in the $MSE$ is the multivariate analysis of variance, followed by the fixed effects model, then the one proposed by \citet{deaton1985panel}, and finally, the one proposed in this document, which adjusts an additional random effect from subgroups of observational units. For a better visualization of these last two models, analogous to Figure \ref{simulation:fig4}, Figure \ref{simulation:fig5} shows the distributions of the $MSE$ from the Deaton model and the proposed one, where it is observed that the last one has lower values.

\begin{figure}
    \includegraphics[width=10cm]{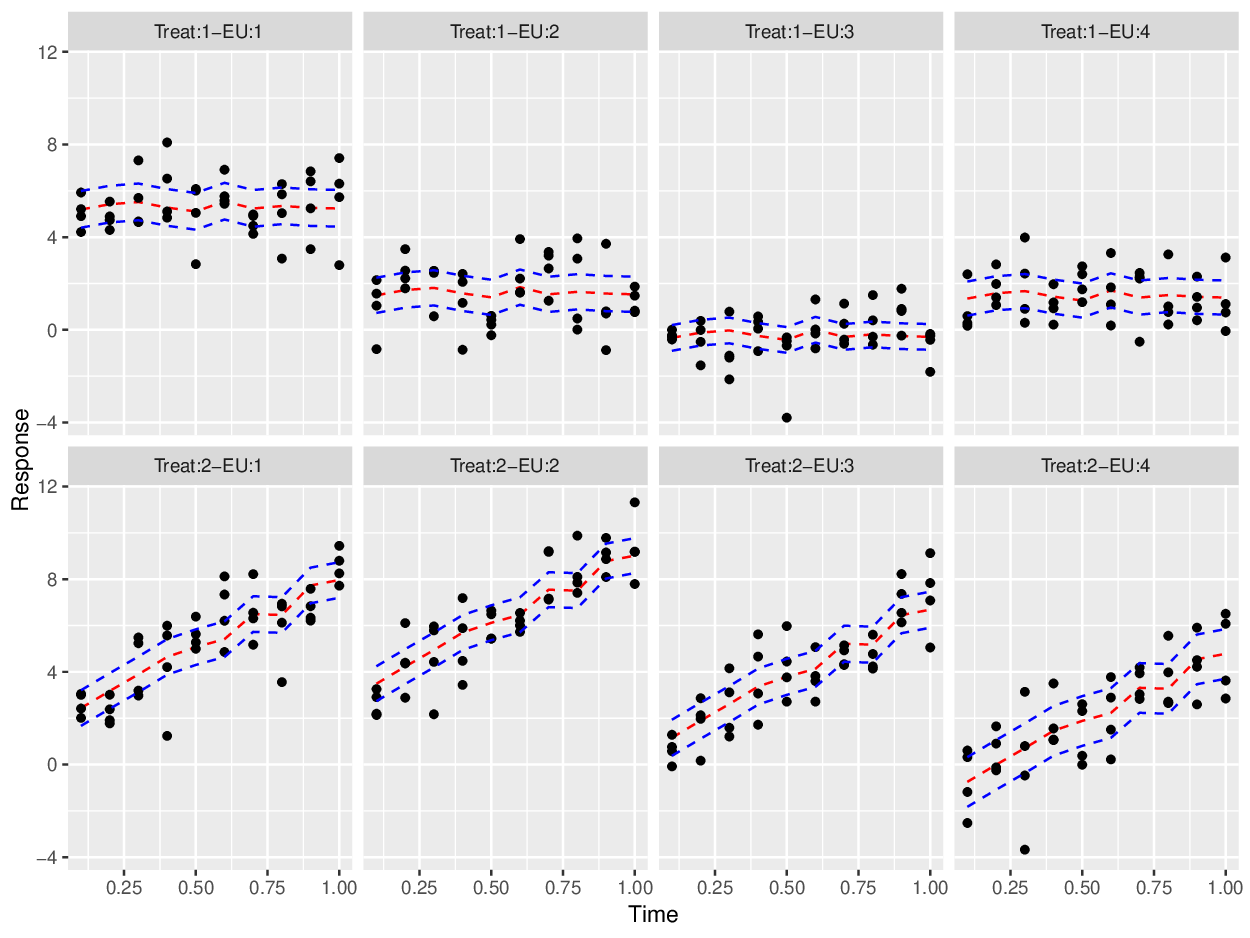}
    \caption{Scatter diagram of the points observed in time together with the estimates made by the average models (red line) and the one proposed (blue line), which considers two subgroups within the experimental units, one of high values and another of low values.}\label{simulation:fig3}
\end{figure}

By way of illustration and analogously to Figure \ref{simulacion:fig1}, Figure \ref{simulation:fig3} shows the observed values (black dots) and those adjusted by the models described in the sections \ref{simulation:sec_deaton} (red line) and \ref{simulation:sec_proposed} (blue lines). It is important to note that the observed values of the proposed model are divided into two groups (one upper and one lower) to estimate a random effect $\eta_{r(i)}$,
trying to make an analogy to the initially simulated $\eta_{j(i)}$ coefficients, and that is, the reason why two blue lines are displayed in such figure. Additionally, it is observed that through the proposed model, a better fit is obtained on the observed points.

\begin{figure}
    \includegraphics[width=10cm]{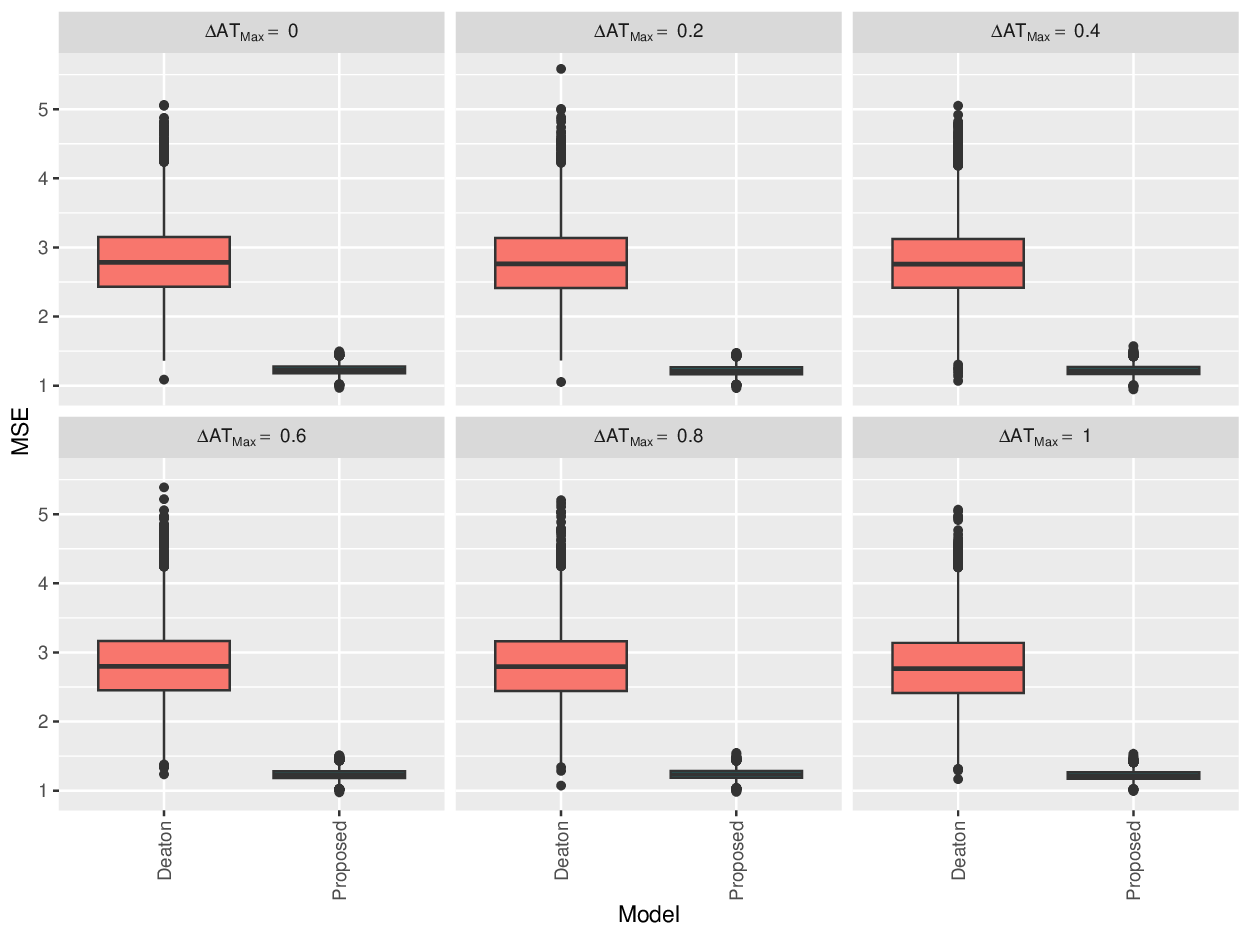}
    \caption{$MSE$ for one hundred simulations of the destructive sampling scheme for the model described in \citet{deaton1985panel} and the one proposed in this paper.} \label{simulation:fig5}
\end{figure}

From the simulated model, it is possible to initially hypothesize significant differences in the coefficients of the interaction $AT_{ij}$, that is, $H_0^{(1)}$, which was previously described. If $H_0^{(1)}$ is not rejected, it is of interest to contrast the hypotheses of significant differences between treatments and between times, that is, to evaluate $H_0^{(2)}$ and $H_0^{(3) }$, respectively \citep{hinkelmann2011design}. It should be taken into account that depending on the conclusion about $H_0^{(1)}$, the contrasts $H_0^{(2)}$ and $H_0^{(3)}$ can be developed.

\begin{figure}
    \includegraphics[width=10cm]{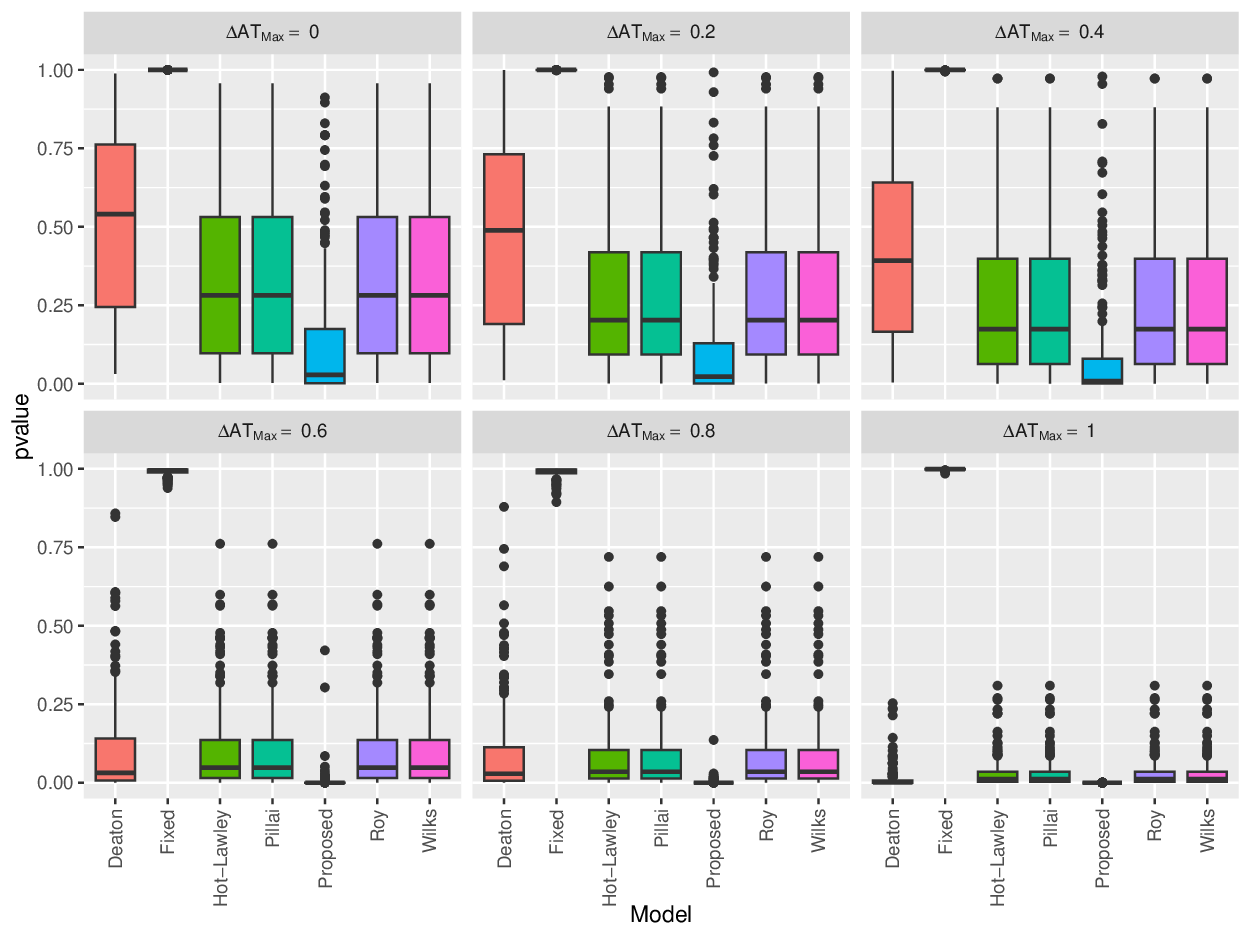}
    \caption{Boxplots for the distributions of the p-values of the interaction significance tests for each of the models and under different maximum differences between the coefficients $AT_{ij}$.}\label{simulation:fig7}
\end{figure}

To contrast $H_0^{(1)}$, there is a p-value from the original model that uses the complete data, which would be ideal, but in practice, it cannot be obtained due to the conditions of the model. Despite this, the coefficient is the one that is closest to reality, and therefore, it will be used as a reference in the following comparison. Considering the four methodologies implemented, their results are compared with the reference p-value obtained from the simulated regression. To do this with each set of complete data, the loss of information was simulated on one hundred occasions, and in this way in each of these repetitions the p-value associated with $H_0^{(1)}$ was collected in each of the models subjected to comparison. 
This procedure was performed for different scenarios that depend on the value assigned to $\Delta AT_{max}$. Figure \ref{simulation:fig7} illustrates this comparison, where each box refers to a $\Delta AT_{max}$. On the other hand, the boxplots show the distributions of the p-values resulting from the different methodologies in each of the scenarios. Note that the four test statistics described in Table \ref{frame:tab2} were taken into account in the multivariate analysis of the variance test.
\begin{table}
  \centering
  \begin{tabular}{c c }
  \hline
  Test Statistics & Formula \\ \hline
  Pillai & $\displaystyle V^{(a)}=\sum _{i=1}^a \dfrac{\lambda_i}{1+\lambda_i}$ \\
  Lawley-Hotelling & $\displaystyle U^{(a)}=\sum _{i=1}^a \lambda_i$ \\
  Wilk's Lambda & $\displaystyle \Lambda=\prod_{i=1}^a \frac{1}{1+\lambda_i}$ \\
  Roy's root & $\displaystyle \theta= \frac{\lambda _{(1)}}{1+\lambda _{(1)}}$ \\ \hline
  \end{tabular}\caption{Formulas for test statistics for multivariate analysis of variance, which are calculated from the eigenvalues of the matrix $\pmb{Q}_e^{-1}\pmb{Q }_h$, denoted by $\lambda_i$, while the highest eigenvalue is denoted by $\lambda_{(1)}$.}\label{frame:tab2}
  \end{table}

In Figure \ref{simulation:fig7}, the values most similar to the reference p-value, in each of the cases, are those associated with the proposed model, which is mainly because it includes an additional random factor. After the proposed model, the tests carried out with the multivariate analysis are the ones that come closest to the line determined by the model with all the observations, followed by the Deaton model and finally, the one that only includes fixed effects. Furthermore, as expected as $\Delta AT_{max}$ (values in the headers) increase, all tests tend to reject $H_0^{(1)}$.

\begin{figure}
    \includegraphics[width=12cm]{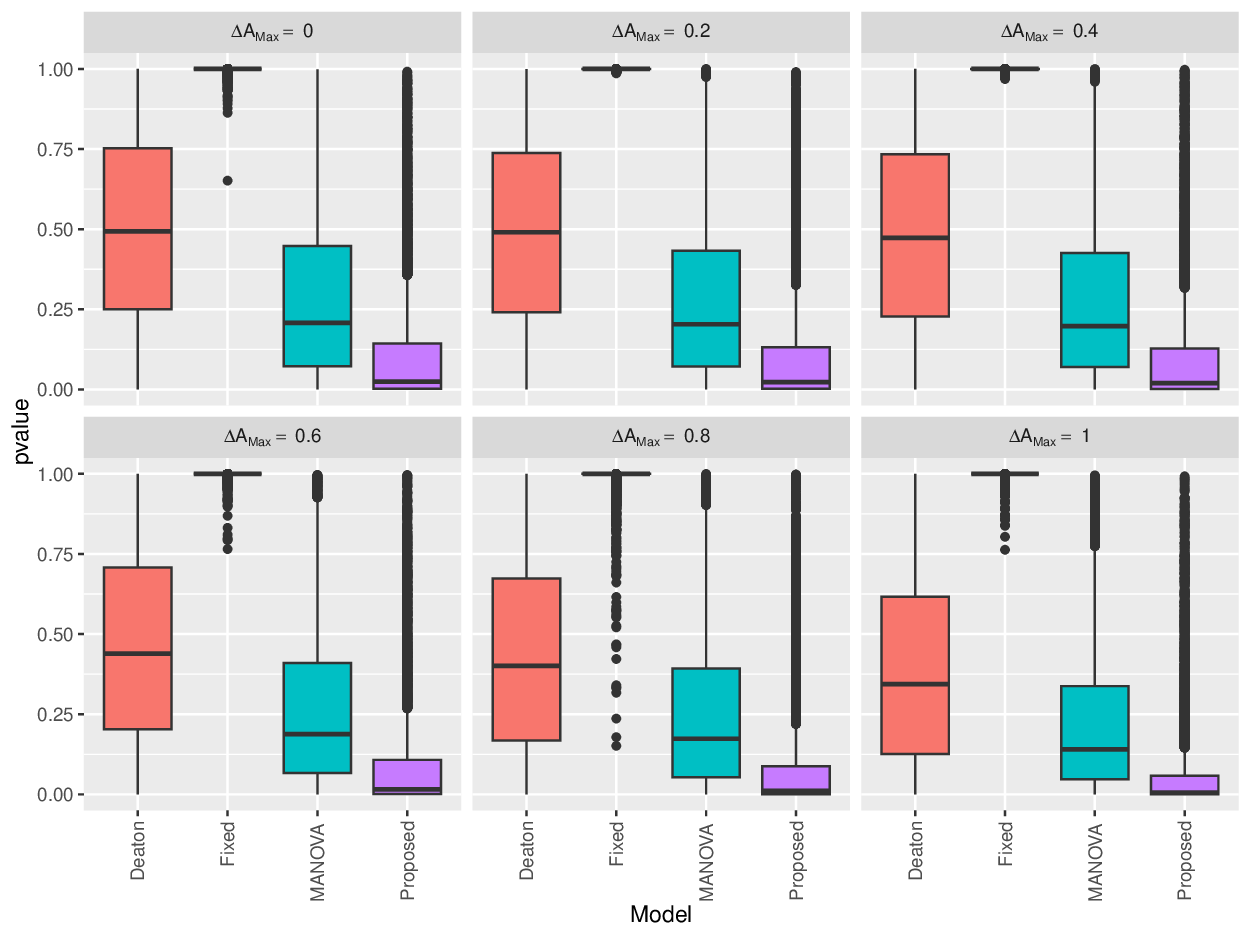}
    \caption{Boxplots for the distributions of the p-values of the tests carried out to contrast $H_0^{(2)}$ for each of the models and under different maximum differences between the coefficients $A_{i }$.}\label{simulation:fig8}
\end{figure}

Similarly to the exercise carried out with the interactions $AT_{ij}$, the same exercise in Figure \ref{simulation:fig8} was carried out, but replicating it for the main effects $A_i$ of the treatments, that is, to test $H_0 ^{(2)}$ in each of the repetitions, where the red line is at the level of the p-value using the model with complete data, called the reference value. In this way, the exercise was done based on different scenarios considering a difference between the two levels of the factor $A$, that is, $\Delta A_{max}$. Boxplots represent the distributions of the p-values in the hundred simulations for each of the coefficients (values in the box headers). In this case, the only model with a p-value that decreases as the parameter values are larger is the proposed model. This means that if the interest is only to detect differences between the effects of the main factor $A_i$, the model \citet{deaton1985panel} and the other models do not show good results, but the proposed one is better in terms of the test statistics and their respective p-values. Following these models, the tests are developed using multivariate analysis of variance.

\begin{figure}
    \includegraphics[width=12cm]{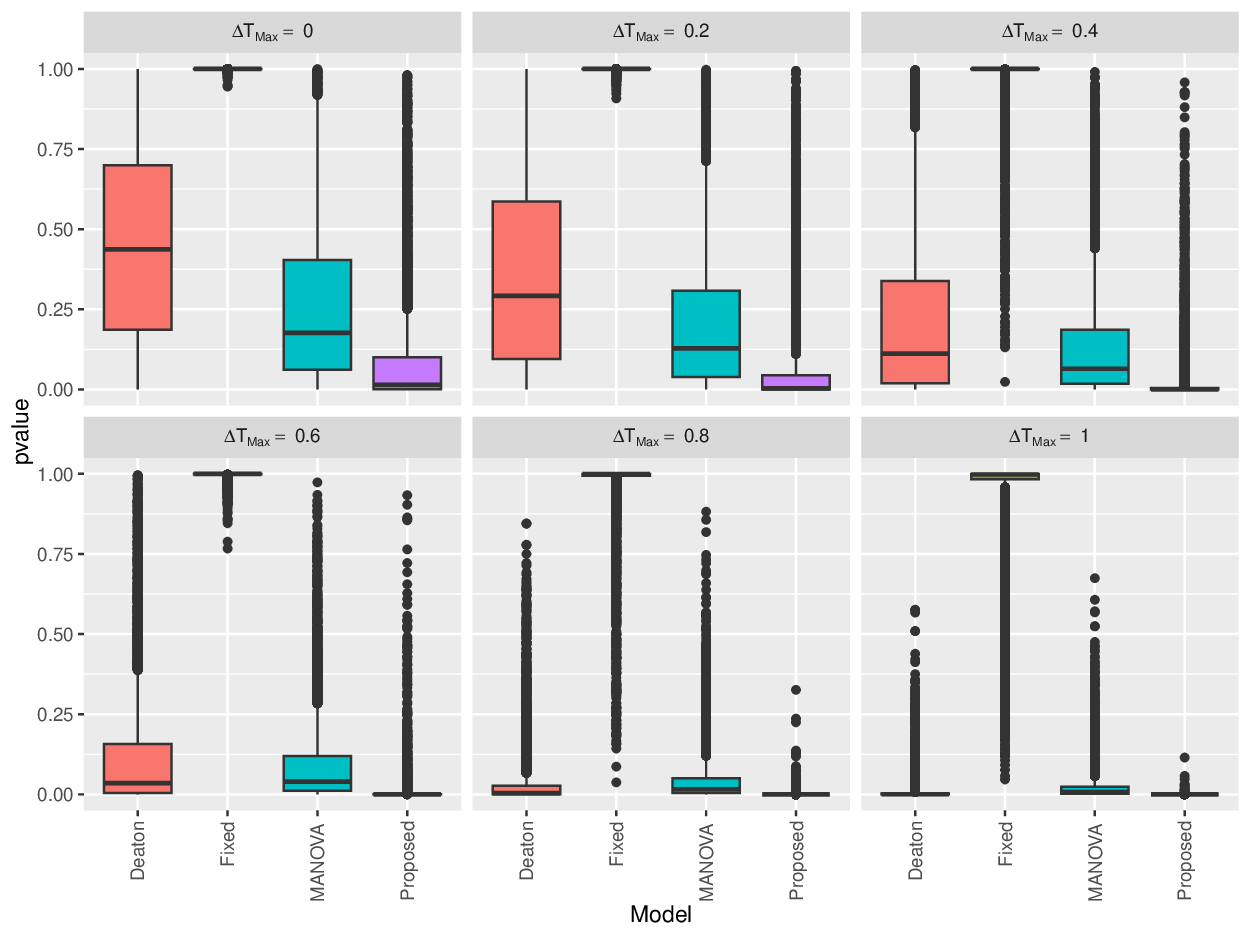}
    \caption{Boxplots for the distributions of the p-values of the tests carried out to contrast $H_0^{(3)}$ for each of the models and under different maximum differences between the coefficients $T_{k }$.}\label{simulacion:fig10}
\end{figure}

Finally, the same exercise was performed to evaluate $H_0^{(3)}$. That is, different values were assigned for the maximum difference between effects associated with the different $T_k$, denoted in this case as $\Delta T_{max}$. The comparison of p-values associated with the four methodologies, together with the reference value calculated from the model that included all the data, is shown in Figure \ref{simulacion:fig10}. The values associated with the average model and the proposed one are those that are closest to the reference p-value.

\begin{figure}
    \includegraphics[width=12cm]{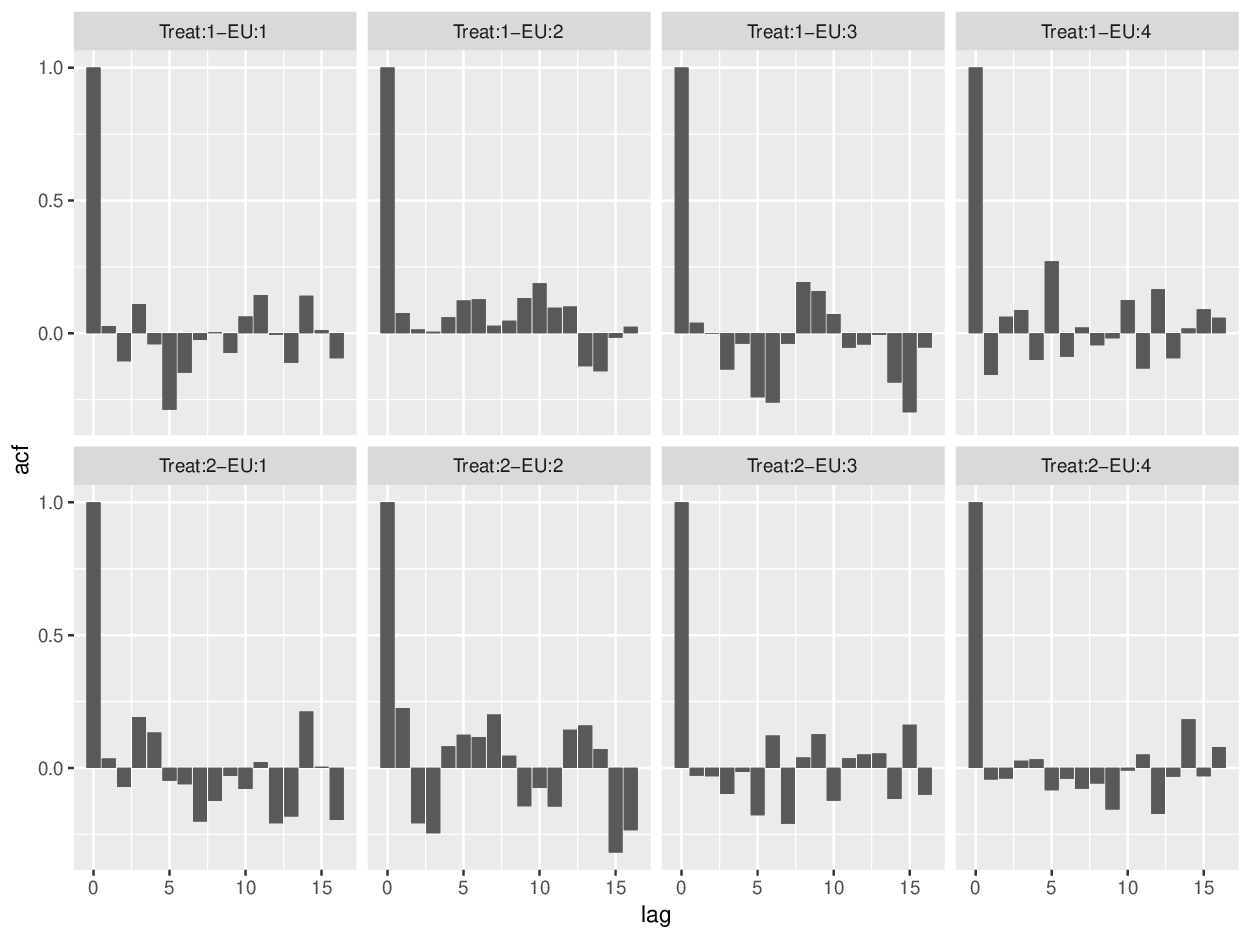}
    \caption{Correlograms for the subgroups of observational units within the different experimental units and for the two factors under consideration in the residuals of one of the model fits given in Equation \eqref{simulacion:eqmodsimula_4}.}\label{simulation:fig9}
\end{figure}

The proposed model comes from the evaluation of the temporal autocorrelation included within the simulated model. That is, based on the model given in Equation \eqref{simulacion:eqmodsimula} where a temporal correlation was included, more specifically a first-order autocorrelated process ($AR(1)$), one can think about evaluating the presence or absence of this after the destructive process of observational units in the final model. In this way, Figure \ref{simulation:fig9} shows the correlograms associated with the residuals of the proposed model, described in Equation \eqref{simulacion:eqmodsimula_4} for the different groups of formed observational units. There is no evidence of temporal autocorrelation. It is recovered through the mixed effects models (proposed and pseudo-panel) with the formed groupings.

\section{Application}

To carry out the application of the considered models, the mathematics results corresponding to the ICFES or Saber 11 tests were taken from schools located in the different municipalities of Colombia, which must be presented by all students at the end of the eleventh grade as a prerequisite for their secondary education degree. As previously explained, the majority of students who once took this test do not take it in subsequent years, and it can be assumed that even though the students of a specific school are not the same in different periods. Their results can be similar since they have similar training. This is because they shared the educational institution, teachers, teaching methodologies, and facilities, among other possible factors that may influence learning. Considering that all institutions are represented by their students in all measurement periods, each school represents an experimental unit, while the students from whom the measurements come are the observational units.

The data was found on the open data page provided by the Ministry of Information and Communications Technologies (see \citet{datossim}).
The mathematics score was taken as the response variable and the following were taken as independent variables in the models: i) location area of the institution (rural or urban), ii) gender of students at the institution (female, male, or mixed), iii) nature of the school (official or private) and, iv)  measurement period.
It is necessary to consider that most schools in Colombia begin their school year in January and end around December, which indicates that these schools present the exam considered in the second semester of the academic calendar. For this reason, only observations were taken whose measurement period corresponds to the second semester, and the years considered range go from 2013 to 2018.

\begin{table}[h]
\centering
\caption{Variance analysis table for the model of Equation \eqref{simulacion:eqmodsimula_3}}
\begin{tabular}{ lrrr }
  \hline
Variable&df& F-statistic &p-value\\ \hline
Location area & 1 & 895.2 & $<0.001$ \\
Gender & 2 & 243.5 & $<0.001$ \\
Nature of school & 1 & 1069.8 & $<0.001$ \\
Period & 1 & 74179.5 & $<0.001$ \\
Gender*Nature & 2 & 9.9 & $<0.001$ \\
Gender*Area & 2 & 1.4 & 0.2549 \\
  \hline
\end{tabular}\label{real_data:table_2}
\end{table}

Considering that the observational units are the students, but not the educational establishments, one can think about including a random effect to discriminate the variability between and within schools. In this way, what is done in the proposed model is to divide the students into those with good performance and those with poor performance based on the median scores in each school. These groups can be monitored over time, with the objective of including a random effect referring to said groups. For this reason, Equation \eqref{real_data:eq_3} shows the model analogous to the one described in Section \ref{simulation:sec_proposed} applied in this context:
\begin{equation}
    \label{real_data:eq_3}
y_{ijklnmr}=\mu+N_i+G_j+A_k+P_l+NG_{ij}+AG_{jk}+b_n+\eta^*_{nm}+\epsilon_{ijklnmr}
\end{equation}

  \noindent $\text{with} \enspace i=1,2,\enspace j=1,2,3, \enspace k=1,2, \enspace l=1,\ldots,6 \enspace \text{and } \enspace m=1,\enspace 2,$ constrained to
\begin{equation}  \sum _{i=1}^2 N_i= \sum _{j=1}^3 G_j=\sum _{k=1}^2 A_k=\sum _{l=1}^6 P_l=
  \sum _{i=1}^2 NG_{ij}=\sum _{j=1}^3 NG_{ij}=\sum _{j=1}^3 AG_{jk}=\sum _{k=1}^ 2 AG_{ij}=0 \nonumber
\end{equation}

where $y_{ijklnmr}$ is the math score of the $r$-th student in the $m$-th group at $n$-th school for the $l $-th period, in the $i$-th type (nature of the school) with the $j$-th group of the gender variable in the $k$-th location area, $b_n$ and $\eta_{r(nm)}$ are the random effects of the $r$-th school and the $m$-th group at the $n$-th school, respectively. The subindex $m$ only takes the values 1 and 2, since the measurements of each institution were divided into two groups (good and bad performance) based on their median, but despite this, the partitions can be made in a way that more groups are formed, through percentiles or quartiles for example. The respective analysis of variance table is shown in Table \ref{real_data:table_3}, where the same is concluded as in Table \ref{real_data:table_2}.

\begin{table}[h]
\centering
\caption{Analysis of variance for the model of Equation \eqref{real_data:eq_3}.}
\begin{tabular}{ lrrr }
  \hline
Variable&df& F-statistic &p-value\\ \hline
Location area & 1 & 427.4 & $<0.001$ \\
Gender & 2 & 203.5 & $<0.001$ \\
Nature of school & 1 & 649.5 & $<0.001$ \\
Period & 1 & 195259.9 & $<0.001$ \\
Gender*Nature & 2 & 9.8 & $<0.001$ \\
Gender*Area & 2 & 1.0 & 0.3865 \\
  \hline
\end{tabular}\label{real_data:table_3}
\end{table}

Finally, for the multivariate analysis of variance model analogous to Section \ref{simulation:sec_manova}, the average of each student group within the institutions was taken for each period. That is, the $\bar{y}_{ijklnm\cdot}$ forms the response vectors, while the equation that considers the fixed effects remains the same as that of the univariate linear models described previously. The results of the MANOVA using the Pillai statistic (considering that the tests gave similar results, the other three were omitted) are shown in Table \ref{real_data:table_4}.

\begin{table}[h]
\centering
\caption{Variance analysis table for the MANOVA whose response is the average scores in mathematics in the groups made up of students within the schools.}
\begin{tabular}{ lrrrrrr }
  \hline
Variable &df &
  {Pillai} &
  {Approx. F} &
  {Num. df} &
  {Den. df} &
 {p-value}\\ \hline

Location area & 1 & 0.06031 & 185 & 6 & 17261 & $<0.00001$\\
Gender & 2 & 0.02624 & 38 & 12 & 34524 & $<0.00001$ \\
Nature of school & 1 & 0.05518 & 168 & 6 & 17261 & $<0.00001$\\
Gender*Nature & 2 & 0.00280 & 4 & 12 & 34524 & $<0.00001$\\
Gender*Area & 2 & 0.00077 & 1 & 12 & 34524 & 0.3425 \\
  \hline
\end{tabular}\label{real_data:table_4}
\end{table}

In other words, results similar to those obtained through those models that included mixed effects were obtained using this last model.

As a complementary comparison exercise, Table \ref{real_data:table_5} is presented, which shows the mean squares error, the correlation between the estimates and the original response, and the latter squared, calculating what we call a $ pseudo-R^2$, to obtain goodness-of-fit measures. There, it is observed that the best fitting model is the one described in Equation \eqref{real_data:eq_3}, which includes two random effects in addition to the errors, that is, the proposed model since it has a mean square of the smallest error compared to the others and higher values in terms of correlation and $pseudo-R^2$.

\begin{table}[ht]
\centering
\caption{Summary of the models applied to the mathematics score data, in terms of the mean square error and the correlation between the estimates and the observed values.}
\begin{tabular}{lrr r}
   \hline{Model} &{$MSE$} &{$r_{y,\hat{y}}$} &{$pseudo-R^2$}\\ \hline
Fixed effects & 123.28 & 0.27 & 0.07 \\
Deaton & 93.18 & 0.55 & 0.30 \\
Proposed & 39.61 & 0.84 & 0.70 \\
MANOVA & 88.04 & 0.38 & 0.14 \\
    \hline
\end{tabular}\label{real_data:table_5}
\end{table}

Considering that the best model is the one corresponding to Equation \eqref{real_data:eq_3}, some validations were carried out on the random effects and the residuals.


To evaluate the assumption of normality, the Anderson-Darling test was performed for the three random effects. The respective p-values of these tests are seen in Table \ref{real_data:table_6}. Since the p-values are greater than 0.05, for the three errors it is concluded that there is not enough statistical evidence to reject the hypothesis of normality in the random errors.

\begin{table}
\centering
\caption{p-values associated with the Anderson-Darling test to test the hypothesis of normality of the random effects included in model \eqref{real_data:eq_3}.}
\begin{tabular}{c c cc}
   \hline
   Test & $\epsilon_{ijklnmr}$ & $\eta_{nm}$ & $b_n$ \\
   \hline
   Anderson-Darling & 0.06 & 0.13 & 0.11 \\
    \hline
\end{tabular}
\label{real_data:table_6}
\end{table}
On the other hand, to evaluate the assumption of homoscedasticity in the variances of the random effects in the proposed model, categories are randomly assigned to the coefficients $b_n$, $\eta^*_{nm}$ and $\epsilon_{ijklnmr}$. In such way, comparing the coefficients across these groupings, it can be determined whether the assumption of constant variances is met. Table \ref{real_data:table_7} shows the p-values associated with the constant variance test for the different random effects included in the proposed model applied to the real data given in Equation \eqref{real_data:eq_3}. In this table, it is observed that for the case of $\epsilon_{ijklnmr}$ the hypothesis of homoscedasticity is rejected, and it is concluded that the variance is not constant; however, in the other two random factors, it is concluded that there is no evidence to conclude that the variance is not constant, using a significance level of 0.05.

\begin{table}
\centering
\caption{p-values associated with the Bartlett test to test the hypothesis of homoscedasticity of the random effects included in model \eqref{real_data:eq_3}.}
\begin{tabular}{c c cc}
   \hline
   Test & $\epsilon_{ijklnmr}$ & $\eta_{nm}$ & $b_n$ \\
   \hline
   Bartlett & 0.02 & 0.9545 & 0.5337\\
    \hline
\end{tabular}
\label{real_data:table_7}
\end{table}

\section{Conclusions}

Considering the sampling scheme worked in this document under the longitudinal data scenario, it is concluded that the simple regression model is the least appropriate since the omission of factors such as those associated with the different experimental units leads to opposite conclusions. The proposed model that makes groupings of observational units within the different experimental units is the one that presents the lowest MSE, for the other three regression models taken into account. This adjustment in incomplete data is the one that is closest to the p-value calculated to evaluate the significance of the main and secondary effects, through a regression model considered without loss of information, followed by the tests developed with MANOVA, in the case of the interaction and the model proposed by \citet{deaton1985panel} to evaluate the main effects.
Additionally, the temporal autocorrelation simulated in the residuals of the original model is not recovered in the formed groupings, which is because it is not the same observational units that are measured at all times. Concerning the application to real data, an important contribution is seen from the variables location of the institution, gender of the students at school, their type, measurement period, and the interaction between gender and nature of the school (private or public). The interaction between gender and location area is a clear example of the risks of omitting factors, since in the exclusive fixed effects model it was significant, while in the other three, it was not. In this last, as in the simulation, the proposed model is the one with the best goodness-of-fit, which was observed through the MSE, the correlation between the adjusted and observed values and the latter statistic squared, exerting a role of a pseudo-R2.
\appendix
\section{Proofs}\label{apenA}
\begin{proof}
The proposed model is defined in Equation \eqref{marco:modmix1}, and the best linear unbiased predictor of $\pmb{y}=(\pmb{y}_1, \ldots, \pmb{y}_n)$ for that linear mixed model is:
\begin{equation}\nonumber
    \hat{\pmb{y}} = \pmb{X}\hat{\pmb{\beta}}+ \hat{\pmb{\Psi}}\pmb{Z}\hat{\pmb{\Sigma}}^{-1}(\pmb{y}-\pmb{X}\hat{\pmb{\beta}})
\end{equation}
Furthermore, the total sum of squares of $\pmb{y}$ is given by the expression:
\begin{equation}\nonumber
\pmb{y}^t\pmb{y} = \hat{\pmb{\beta}}^t\pmb{R_X}^t\pmb{R_X} \hat{\pmb{\beta}}^t +  \hat{\pmb{\beta}}^t\pmb{R_{ZX}}^t\pmb{R_{ZX}} \hat{\pmb{\beta}}^t +  \hat{\pmb{v}}^t\pmb{R_{ZZ}}^t\pmb{R_{ZZ}} \hat{\pmb{v}}^t + SS^{(0)}_{Resid}
\end{equation}
where the matrices $\pmb{R_{ZZ}}$, $\pmb{R_{ZX}}$ and $\pmb{R_{X}}$ are obtained by the following Cholesky decomposition \cite{bates2015}:
\begin{align}
   \pmb{R_{X}}^t \pmb{R_{X}} &= \pmb{{X}}^t \pmb{{X}} -\pmb{R_{ZX}}^t \pmb{R_{ZX}}\nonumber\\
   \pmb{\Delta}^t \pmb{Z}^t\pmb{X} &= \pmb{R_{ZZ}}\pmb{R_{ZX}}\nonumber\\
   \pmb{\Delta}^t \pmb{Z}^t \pmb{Z} \pmb{\Delta}+ \mathbf{I}&=\pmb{R_{ZZ}}^t\pmb{R_{ZZ}}\nonumber 
\end{align}
and since the matrix $\pmb{Z}$ contains both the indicator variables of the random effects of each experimental unit $b_i$ and the random effects of each subgroup $\eta_{1(i)}, \ldots, \eta_{G_i(i)}$, it can be written $\pmb{Z}=[\pmb{Z_b}\vdots \pmb{Z_\eta}]$
Now, for the model defined in Equation \eqref{marco:modmix12} we have:
\begin{equation}\nonumber
\pmb{y}^t\pmb{y} = \hat{\pmb{\beta}}^t\pmb{R_X}^t\pmb{R_X} \hat{\pmb{\beta}}^t +  \hat{\pmb{\beta}}^t\pmb{R_{Z_{b}X}}^t\pmb{R_{Z_bX}} \hat{\pmb{\beta}}^t +  \hat{\pmb{b}}^t\pmb{R_{Z_{b}Z_{b}}}^t\pmb{R_{Z_{b}}} \hat{\pmb{b}}^t + SS^{(iii)}_{Resid}
\end{equation}
then
\begin{equation}\nonumber
    \hat{\pmb{b}}^t\pmb{R_{Z_{b}Z_{b}}}^t\pmb{R_{Z_{b}}} \leq \hat{\pmb{v}}^t\pmb{R_{ZZ}}^t\pmb{R_{ZZ}} \hat{\pmb{v}}^t
\end{equation}
Therefore $SS^{(0)}_{Resid}< SS^{(iii)}_{Resid}$.  Analogously, we have that for the estimation of the model defined in equation \eqref{marco:modmix11}:
\begin{equation}\nonumber
\pmb{y}^t\pmb{y} = \hat{\pmb{\beta}}^t\pmb{R_X}^t\pmb{R_X} \hat{\pmb{\beta}}^t + SS^{(ii)}_{Resid}
\end{equation}
Then, $SS^{(iii)}_{Resid}< SS^{(ii)}_{Resid}$. Now, since the variables that conform $\pmb{X}$ are all binary, the estimators of $\pmb{\beta}$ are the same in the 4 estimation models. However, since in Equation \eqref{marco:modmix10} the average of all the observations of the experimental unit is performed for each time, it can be stated that:
\begin{align*}
\pmb{y}^t\pmb{y} & = \sum_{ij}(y_{ijk}-\bar{y}_{i\cdot k})^2 + \hat{\pmb{\beta}}^t\pmb{R_{X}}^t\pmb{R_{X}} \hat{\pmb{\beta}}^t +\\
&\hat{\pmb{\beta}}^t\pmb{R_{\bar{Z}_{b}{X}}}^t\pmb{R_{\bar{Z}_b{X}}} \hat{\pmb{\beta}}^t +  \hat{\pmb{b}}^t\pmb{R_{\bar{Z}_{b}\bar{Z}_{b}}}^t\pmb{R_{\bar{Z}_{b}}} \hat{\pmb{b}}^t + SS^{(iii)}_{Resid}
\end{align*}
where $\pmb{\bar{Z}_{b}}$ is constructed with the random effects of each experimental unit, but only the average of each experimental unit per period is used, that is, it has the same columns as $\pmb{{{Z}_{b}}}$, but fewer rows. Therefore $$SS^{(ii)}_{Resid}=SS^{(iii)}_{Resid}+\sum_{ij}(y_{ijk}-\bar{y}_{i\cdot k})^2$$ so, $ SS^{(iii)}_{Resid}\leq SS^{(i)}_{Resid}\leq SS^{(ii)}_{Resid}$. 
\end{proof}
\section{Supplementary information}
Supplementary file 1:{R codes for the estimation of model and simulation study presented in the paper.}\label{sf1}
\bibliographystyle{statistica}

\begin{thebibliography}{29}
\providecommand{\natexlab}[1]{#1}
\providecommand{\url}[1]{\texttt{#1}}
\providecommand{\urlprefix}{URL }

\bibitem[{Ariyo \emph{et~al.}(2020)Ariyo, Quintero, Mu{\~n}oz, Verbeke, and Lesaffre}]{ariyo2020bayesian}
\textsc{O.~Ariyo}, \textsc{A.~Quintero}, \textsc{J.~Mu{\~n}oz}, \textsc{G.~Verbeke}, \textsc{E.~Lesaffre} (2020).
\newblock \emph{Bayesian model selection in linear mixed models for longitudinal data}.
\newblock Journal of Applied Statistics, 47, no.~5, pp. 890--913.

\bibitem[{Bates \emph{et~al.}(2015{\natexlab{a}})Bates, M{\"a}chler, Bolker, and Walker}]{lme4}
\textsc{D.~Bates}, \textsc{M.~M{\"a}chler}, \textsc{B.~Bolker}, \textsc{S.~Walker} (2015{\natexlab{a}}).
\newblock \emph{Fitting linear mixed-effects models using {lme4}}.
\newblock Journal of Statistical Software, 67, no.~1, pp. 1--48.

\bibitem[{Bates \emph{et~al.}(2015{\natexlab{b}})Bates, Mächler, Bolker, and Walker}]{bates2015}
\textsc{D.~Bates}, \textsc{M.~Mächler}, \textsc{B.~Bolker}, \textsc{S.~Walker} (2015{\natexlab{b}}).
\newblock \emph{Fitting linear mixed-effects models using lme4}.
\newblock Journal of Statistical Software, 67, no.~1, p. 1–48.
\newblock \urlprefix\url{https://www.jstatsoft.org/index.php/jss/article/view/v067i01}.

\bibitem[{Canavire-Bacarreza and Robles(2017)}]{canavire2017non}
\textsc{G.~Canavire-Bacarreza}, \textsc{M.~Robles} (2017).
\newblock \emph{Non-parametric analysis of poverty duration using repeated cross section: an application for {Peru}}.
\newblock Applied Economics, 49, no.~22, pp. 2141--2152.

\bibitem[{Covacevich \emph{et~al.}(2021)Covacevich, Mann, Santos, and Champaud}]{covacevich2021indicators}
\textsc{C.~Covacevich}, \textsc{A.~Mann}, \textsc{C.~Santos}, \textsc{J.~Champaud} (2021).
\newblock \emph{Indicators of teenage career readiness: An analysis of longitudinal data from eight countries}.
\newblock OECD.

\bibitem[{Deaton(1985)}]{deaton1985panel}
\textsc{A.~Deaton} (1985).
\newblock \emph{Panel data from time series of cross-sections}.
\newblock Journal of {Econometrics}, 30, no. 1-2, pp. 109--126.

\bibitem[{Federer and King(2007)}]{federer2007variations}
\textsc{W.~T. Federer}, \textsc{F.~King} (2007).
\newblock \emph{Variations on split plot and split block experiment designs}, vol. 654.
\newblock John Wiley and Sons, Hoboken.

\bibitem[{Finch \emph{et~al.}(2016)Finch, Bolin, and Kelley}]{finch2016multilevel}
\textsc{W.~H. Finch}, \textsc{J.~E. Bolin}, \textsc{K.~Kelley} (2016).
\newblock \emph{Multilevel modeling using R}.
\newblock Chapman and Hall/CRC, Boca Raton.

\bibitem[{Gardes \emph{et~al.}(2005)Gardes, Duncan, Gaubert, Gurgand, and Starzec}]{gardes2005panel}
\textsc{F.~Gardes}, \textsc{G.~J. Duncan}, \textsc{P.~Gaubert}, \textsc{M.~Gurgand}, \textsc{C.~Starzec} (2005).
\newblock \emph{Panel and pseudo-panel estimation of cross-sectional and time series elasticities of food consumption: The case of {US} and polish data}.
\newblock Journal of Business and Economic Statistics, 23, no.~2, pp. 242--253.

\bibitem[{Himaz and Aturupane(2016)}]{himaz2016returns}
\textsc{R.~Himaz}, \textsc{H.~Aturupane} (2016).
\newblock \emph{Returns to education in {Sri} {Lanka}: a pseudo-panel approach}.
\newblock Education Economics, 24, no.~3, pp. 300--311.

\bibitem[{Hinkelmann(2011)}]{hinkelmann2011design}
\textsc{K.~Hinkelmann} (2011).
\newblock \emph{Design and analysis of experiments, special designs and applications}, vol.~3.
\newblock John Wiley and Sons, Blacksburg.

\bibitem[{Melo \emph{et~al.}(2007)Melo, L{\'o}pez, and Melo}]{Melo}
\textsc{O.~Melo}, \textsc{L.~L{\'o}pez}, \textsc{S.~Melo} (2007).
\newblock \emph{Dise{\~n}o de experimentos: m{\'e}todos y aplicaciones}.
\newblock Editorial Universidad Nacional de Colombia. Bogot{\'a}.

\bibitem[{MinTic(2020)}]{datossim}
\textsc{MinTic} (2020).
\newblock \emph{Datos abiertos {Colombia}}.
\newblock url{https://www.datos.gov.co}.
\newblock Accedido 01-08-2019.

\bibitem[{Pinheiro and Bates(2006)}]{bates2000mixed}
\textsc{J.~Pinheiro}, \textsc{D.~Bates} (2006).
\newblock \emph{Mixed-effects models in S and S-PLUS}.
\newblock Springer Science \& Business Media, New York.

\bibitem[{Pinheiro \emph{et~al.}(2018)Pinheiro, Bates, DebRoy, Sarkar, and {R Core Team}}]{nlme}
\textsc{J.~Pinheiro}, \textsc{D.~Bates}, \textsc{S.~DebRoy}, \textsc{D.~Sarkar}, \textsc{{R Core Team}} (2018).
\newblock \emph{{nlme}: Linear and Nonlinear Mixed Effects Models}.
\newblock \urlprefix\url{https://CRAN.R-project.org/package=nlme}.
\newblock R package version 3.1-131.

\bibitem[{Quintana \emph{et~al.}(2016)Quintana, Johnson, Waetjen, and B.~Gold}]{quintana2016bayesian}
\textsc{F.~A. Quintana}, \textsc{W.~O. Johnson}, \textsc{L.~E. Waetjen}, \textsc{E.~B.~Gold} (2016).
\newblock \emph{Bayesian nonparametric longitudinal data analysis}.
\newblock Journal of the American Statistical Association, 111, no. 515, pp. 1168--1181.

\bibitem[{{R Core Team}(2024)}]{R}
\textsc{{R Core Team}} (2024).
\newblock \emph{R: A Language and Environment for Statistical Computing}.
\newblock R Foundation for Statistical Computing, Vienna, Austria.
\newblock \urlprefix\url{https://www.R-project.org/}.

\bibitem[{Schabenberger and Gotway(2017)}]{schabenberger2017statistical}
\textsc{O.~Schabenberger}, \textsc{C.~A. Gotway} (2017).
\newblock \emph{Statistical methods for spatial data analysis}.
\newblock Chapman and Hall/CRC, {Boca} {Raton}.

\bibitem[{Shi \emph{et~al.}(2021)Shi, Dong, Wang, and Cao}]{shi2021functional}
\textsc{H.~Shi}, \textsc{J.~Dong}, \textsc{L.~Wang}, \textsc{J.~Cao} (2021).
\newblock \emph{Functional principal component analysis for longitudinal data with informative dropout}.
\newblock Statistics in Medicine, 40, no.~3, pp. 712--724.

\bibitem[{Sprietsma(2012)}]{sprietsma2012computers}
\textsc{M.~Sprietsma} (2012).
\newblock \emph{Computers as pedagogical tools in {Brazil}: a pseudo-panel analysis}.
\newblock Education Economics, 20, no.~1, pp. 19--32.

\bibitem[{Tovar \emph{et~al.}(2012)Tovar, Zulaica, and N{\'u}{\~n}ez-Ant{\'o}n}]{tovar2012analysis}
\textsc{A.~O. Tovar}, \textsc{I.~G. Zulaica}, \textsc{V.~N{\'u}{\~n}ez-Ant{\'o}n} (2012).
\newblock \emph{Analysis of pseudo-panel data with dependent samples}.
\newblock Journal of Applied Statistics, 39, no.~9, pp. 1921--1937.

\bibitem[{UNESCO(2012)}]{unesco2012international}
\textsc{UNESCO} (2012).
\newblock \emph{International standard classification of education: Isced 2011}.
\newblock Comparative Social Research, 30.

\bibitem[{Urdinola and Ospino(2015)}]{urdinola2015long}
\textsc{B.~P. Urdinola}, \textsc{C.~Ospino} (2015).
\newblock \emph{Long-term consequences of adolescent fertility: The colombian case}.
\newblock Demographic Research, 32, pp. 1487--1518.

\bibitem[{Verbeek(2008)}]{verbeek2008pseudo}
\textsc{M.~Verbeek} (2008).
\newblock \emph{Pseudo-panels and repeated cross-sections}.
\newblock In \emph{The econometrics of panel data}, Springer, pp. 369--383.

\bibitem[{Verbeek and Nijman(1993)}]{verbeek1993minimum}
\textsc{M.~Verbeek}, \textsc{T.~Nijman} (1993).
\newblock \emph{Minimum mse estimation of a regression model with fixed effects from a series of cross-sections}.
\newblock Journal of Econometrics, 59, no. 1-2, pp. 125--136.

\bibitem[{Wan \emph{et~al.}(2023)Wan, Zhong, Zhang, and Zou}]{wan2023multikink}
\textsc{C.~Wan}, \textsc{W.~Zhong}, \textsc{W.~Zhang}, \textsc{C.~Zou} (2023).
\newblock \emph{Multikink quantile regression for longitudinal data with application to progesterone data analysis}.
\newblock Biometrics, 79, no.~2, pp. 747--760.

\bibitem[{Wei(2006)}]{wei2006time}
\textsc{W.~W. Wei} (2006).
\newblock \emph{Time series analysis}.
\newblock In \emph{The Oxford Handbook of Quantitative Methods in Psychology: Vol. 2}.

\bibitem[{West \emph{et~al.}(2014)West, Welch, and Galecki}]{west2014linear}
\textsc{B.~T. West}, \textsc{K.~B. Welch}, \textsc{A.~T. Galecki} (2014).
\newblock \emph{Linear mixed models: a practical guide using {Statistical} {Software}}.
\newblock {CRC} Press, Boca Raton.

\bibitem[{Zou \emph{et~al.}(2023)Zou, Zeng, Xiao, and Luo}]{zou2023bayesian}
\textsc{H.~Zou}, \textsc{D.~Zeng}, \textsc{L.~Xiao}, \textsc{S.~Luo} (2023).
\newblock \emph{Bayesian inference and dynamic prediction for multivariate longitudinal and survival data}.
\newblock The Annals of Applied Statistics, 17, no.~3, pp. 2574--2595.

\end{thebibliography}


\begin{abstract}
  This paper proposes an analysis methodology for the case where there is longitudinal data with destructive sampling of observational units, which come from experimental units that are measured at all times of the analysis.
A mixed linear model is proposed and compared with regression models with fixed and mixed effects, among which is a similar that is used for data called pseudo-panel, and one of multivariate analysis of variance, which are common in statistics. To compare the models, the mean square error was used, demonstrating the advantage of the proposed methodology. In addition, an application was made to real-life data that refers to the scores in the Saber 11 tests applied to students in Colombia to see the advantage of using this methodology in practical scenarios.
\end{abstract}

\end{document}